\documentclass[aps,pra,twocolumn,showpacs]{revtex4-1}
\usepackage{graphicx}
\usepackage{amssymb,amsfonts,amsmath}
\usepackage[colorlinks=true,citecolor=Cerulean,linkcolor=RubineRed,urlcolor=Cerulean]{hyperref}
\hypersetup{breaklinks=true}
\usepackage{graphicx}
\usepackage{color}
\usepackage[usenames,dvipsnames]{xcolor}
\usepackage{epstopdf}
\usepackage[normalem]{ulem}

\newcommand{\Heff}{H_\mathrm{eff}}
\newcommand{\Jeff}{J_\mathrm{eff}}
\newcommand{\Jac}{J_\mathrm{ac}}
\newcommand{\im}{\mathrm{i}}
\newcommand{\ac}{\mathrm{ac}}
\renewcommand{\d}{\mathrm{d}}

\begin{document}

\title{Generation of atypical hopping and interactions by kinetic driving}

\author{G.~Pieplow}
\affiliation{Departamento de F\'isica de Materiales, Universidad
Complutense de Madrid, E-28040 Madrid, Spain}

\author{F.~Sols}
\affiliation{Departamento de F\'isica de Materiales, Universidad
Complutense de Madrid, E-28040 Madrid, Spain}

\author{C.E.~Creffield}
\affiliation{Departamento de F\'isica de Materiales, Universidad
Complutense de Madrid, E-28040 Madrid, Spain}

\date{\today}

\begin{abstract}
We study the effect of time-periodically varying the hopping amplitude in a
one-dimensional Bose-Hubbard model, such that its time-averaged value 
is zero. Employing Floquet theory, we derive a static effective Hamiltonian 
in which nearest-neighbor single-particle hopping processes are suppressed, 
but all even higher-order processes are allowed. Unusual many-body features
 arise from the combined effect of nonlocal interactions and correlated 
tunneling. At a critical value of the driving, the system passes from a 
Mott insulator to a superfluid formed by two quasi-condensates with opposite 
nonzero momenta. This work shows how driving of the hopping energy provides a novel form of Floquet engineering, which enables atypical Hamiltonians 
and exotic states of matter to be produced and controlled.
\end{abstract}
\maketitle

\section{Introduction}

Periodically driving a quantum system provides a convenient tool to manipulate
and control its properties by ``Floquet engineering''. In this
technique, Floquet theory is used to describe the effects of a high-frequency 
driving in terms of an effective static Hamiltonian \cite{eckardt_review}. 
This method gives a high degree of control
over the parameters of the effective Hamiltonian, which allows 
these systems to be used as both quantum simulators 
\cite{feynman1982simulating,jaksch,lewenstein2007ultracold,georgescu2014quantum} 
and to treat mathematical problems \cite{creffield2015finding}. 
It is also possible for
the effective Hamiltonian to have properties very different
from the original model, which enables
the realization of systems with exotic properties which cannot be produced 
in other ways, such as Floquet topological insulators 
\cite{Iomin2000,lindner2011floquet} and time crystals 
\cite{time_crystals,time_crystals2}.

In principle, any term of the Hamiltonian can be periodically varied to yield an effective 
model. The earliest forms of driving consisted of applying an 
external potential that oscillated periodically in time. This has the effect of
modifying the tunneling dynamics of the system \cite{dunlap,cdt}, since the 
tunneling terms do not commute with the potential. Manipulating the 
effective tunneling in this way has been used, for example, 
to produce localization \cite{lignier2007dynamical,kierig2008single} and
to drive the Mott transition \cite{eckardt,cec_tsm,zenesini2009parametric},
by setting the effective hopping energy to zero. This form of control
has also been used to produce artificial magnetic fluxes by inducing
phases on the hopping terms, and tuning them to
 mimic the required Peierls phases 
\cite{kolovsky2011creating,creffield2013comment,struck2012tunable,struck2}. 
An alternative method of driving, considered more recently is, to oscillate the 
interaction term with time 
\cite{santos,liberto,gaul2011stability,meinert2016floquet}. 
This again produces a modification of the tunneling terms in the
system, with the unusual feature that the effective (nearest-neighbor) 
hopping depends on the occupation of the sites involved (so-called 
``correlated hopping'' or ``bond-charge interaction'' \cite{nonstandard}). 
Correlated hopping models of this type can also be produced by
applying a resonant driving potential, which gives
the additional possibility of simulating anyon physics \cite{anyon}
by engineering the appropriate occupation-dependent Peierls phases on the 
tunneling elements.

If we consider a Hamiltonian to be composed of potential, interaction, and kinetic terms,
driving of the first two terms has thus already been intensively studied.
In this work we consider the remaining possibility: 
to drive the {\em kinetic} term, which for a lattice system corresponds
to oscillating the {\em tunneling}. 
While the effect of varying the tunneling with time has been studied
previously in cold atom systems \cite{stoeferl,kollath2006spectroscopy,dirks2014theoretical, cardarelli2016} in the context of lattice
modulation spectroscopy, the oscillation is usually taken to be of small amplitude
around a larger constant value. Here, in contrast, we consider the nearest-neighbor hopping 
to oscillate between positive and negative values with a zero time-average.
Although this form of driving may seem rather unusual we discuss 
possible means of achieving it in Section \ref{sec_expt}.
This ``kinetic driving''
produces an unusual effective Hamiltonian in which there is no sharp
distinction between hopping and interaction terms, and where long-range 
correlated and assisted hopping processes occur. 
We consider a one-dimensional system, and show that for small values 
of the driving parameter the dynamics is frozen: 
the system forms a Mott insulator.
As the driving is increased, however, states containing bound doublon-hole pairs, which we term ``dipoles", begin to play an important role. Eventually the system undergoes a quantum phase transition \cite{sachdev} at
which the Mott gap vanishes, and forms
a many-body cat state, consisting of the superposition of two quasi-condensates with opposite, non-zero momenta.

\section{Model}

We consider a one-dimensional Bose-Hubbard (BH) model with $N$ sites
\begin{equation}
H(t) = \sum_{x = 0}^{N-1}\left[\frac{U}{2} n_x(n_x-1) -J(t) \left( a^\dagger_{x+1} a_x + \mathrm{H.c.} \right)\right]  ~,
\label{eq:driven_hubb_ham}
\end{equation}
where $a_x (a_x^\dagger)$ are the usual bosonic annihilation (creation)
operators, $n_x$ is the number operator, and $U$ is the Hubbard
interaction term. The tunneling amplitude between nearest-neighbor 
sites, $J(t)$, is taken to be a $T$-periodic function with zero time-average, 
and we use the specific form $J(t) = \Jac \cos (\omega t)$ 
\footnote{We have also considered squarewave and triangular driving
functions, and find that they give very similar behaviour.}. 
We will take $\hbar = 1$, and for convenience we characterize the amplitude 
of the driving by the dimensionless parameter $\kappa = \Jac / \omega$.
The majority of our results are obtained
for commensurate filling, with the number of bosons equal to $N$. This 
gives a well-defined Mott state in the limit $\kappa \rightarrow 0$. 

As $H(t)$ is time dependent it does not have a set of static eigenvectors 
with associated energies. However, its time-periodicity means that it can 
instead be described in terms of Floquet functions
and quasienergies, which solve the Floquet equation 
$( i \partial_t - H(t) ) | \psi_j(t) \rangle = \epsilon_j | \psi_j(t) \rangle$.
To obtain the quasienergy spectrum for a given
value of $\kappa$, we evaluate the time evolution
operator for one period, $U(T,0)$, and diagonalise it to
obtain its eigenvalues $\lambda_j$. These allow
us to then calculate the quasienergies from the relation
$\lambda_j = \exp (-i T \epsilon_j)$.
The quasienergies are only defined up to integer multiples
of the driving frequency, since $\epsilon_j$ and $\epsilon_j + n \omega$
are not distinguishable, which produces a Brillouin zone structure  
in the quasienergy coordinate \cite{holthaus_review}.
We choose to set $n = 0$, fixing the quasienergies to lie in the 
``first Brillouin zone'', $[0, \omega )$.
As we consider the case of high-frequency driving, $\omega$ is
so large (and the Brillouin zone is thus so wide)
that the quasienergies do not wrap around the zone edges,
and so there is no ambiguity about the quasienergy ordering.
Consequently we can simply treat the quasienergy spectrum within this 
Brillouin zone like a standard energy spectrum, and
in particular we can consider
the lowest quasienergy to be that of the system's ground state.
We shall see later in Section \ref{sec_eff_model} that the spectrum
of the effective Hamiltonian that we derive, which is naturally
well-ordered and possesses a well-defined ground state, indeed reproduces
the quasienergy spectrum, confirming that this interpretation
is justified.

\begin{figure}
\begin{center}
\includegraphics[width=0.45\textwidth,clip=true]{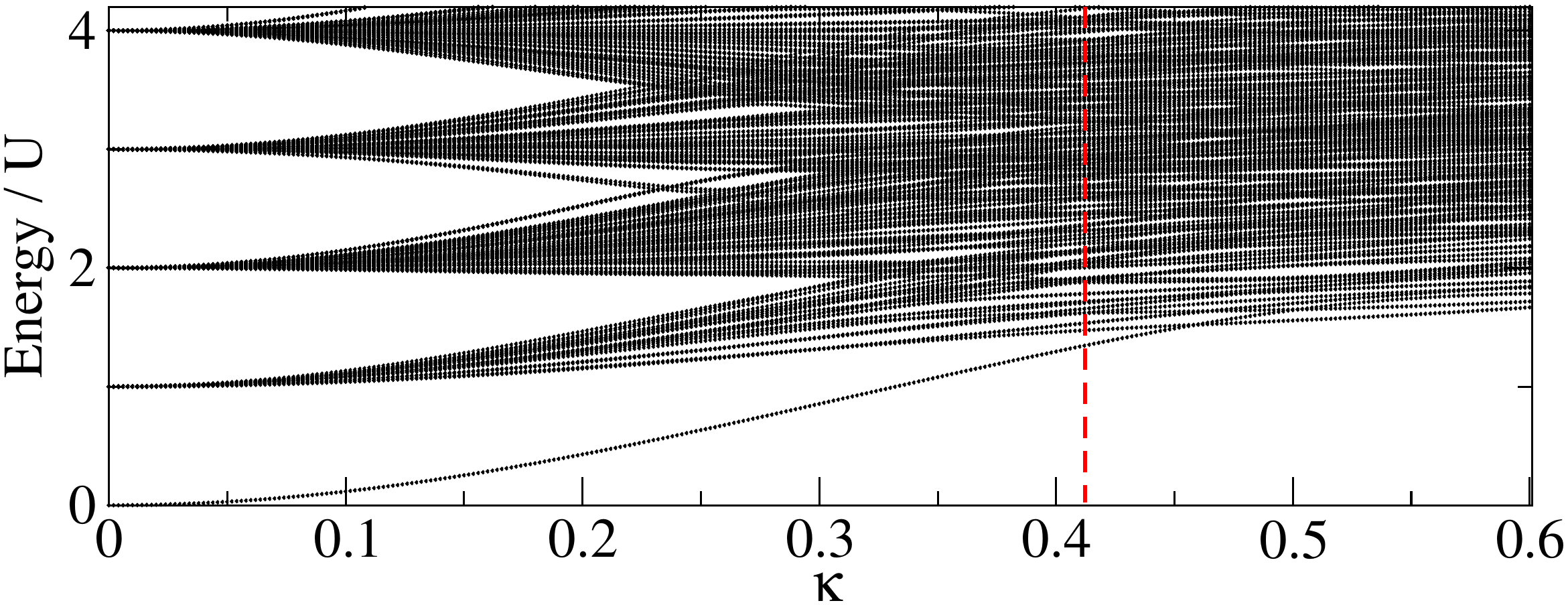}
\end{center}
\caption{Lowest quasienergies for a 6-site model holding 6 bosons.
For small values of $\kappa$ the quasienergies group into well-defined
flat bands separated by the interaction energy $U$. 
As $\kappa$ increases the bands fan out and begin to mix.
The vertical dashed line marks the phase transition between the
Mott insulator and the superfluid, estimated from 
the vanishing of the Mott gap, $\Delta_\infty$, as 
extrapolated to the thermodynamic limit (see Fig. \ref{fig_gap}).} 
\label{spectrum}
\end{figure}

In Fig. \ref{spectrum} we show the quasienergy spectrum obtained for a
6-site, 6-particle model, driven at a high frequency of $\omega = 250 U$.
For $\kappa = 0$ the system is equivalent to a standard BH model with 
$J = 0$. The ground state is thus a perfect Mott state
with each site having unit occupation, while the excited states fall
into separate flat bands that can be
classified by the number and type of multiple site occupations. 
For example, the set of next highest states have one hole and one doublon 
(and so have an energy of $U$), the next set has two holes and
two doublons, and so on. As $\kappa$ is increased from zero, the degeneracies
between the excited states are broken, and the bands begin to fan out and
overlap. Eventually for $\kappa > 0.42$ the ground state approaches the excited states, but little other structure, such as crossings or avoided crossings,
is visible.

\section{\label{sec_eff_model} Effective model}

As shown in detail in Appendix {\ref{app_eff}} 
and discussed, for example, in Refs. \cite{goldman_prx,creffield2016realization}, 
the procedure for obtaining an effective Hamiltonian
consists of Fourier transforming the creation and annihilation
operators, making a unitary transformation to the interaction picture, and then
averaging the transformed Hamiltonian over one period of the driving. The 
result, transformed back to real space, 
constitutes the lowest-order term in the Magnus expansion 
\cite{bukov,anisimovas},
and consists of a sum of 4-operator terms
\begin{equation}
\Heff=\sum_{w,x,y,z=0}^{N-1} Q_{wxyz}  a^\dagger_w a^\dagger_x a_y a_z~,
\label{eq:ham_eff}
\end{equation}
where the transition amplitudes $Q_{wxyz}$ are given by 
\begin{eqnarray}
Q_{wxyz} =& \frac{U}{2N^3}\sum_{l,m,n=0}^{N-1} e^{\im [k_l(x-y)+k_m(x-z)-k_n(x-w)]}
\nonumber \\
& \qquad  M( \kappa; k_l,k_m,k_n) ~,
\label{Q-wxyz}
\end{eqnarray}
and 
\begin{eqnarray}
M( \kappa; k_l,k_m,k_n) &=& \mathcal{J}_0 \Big\{
2 \kappa \big[\cos(k_l)+\cos(k_m) \nonumber \\
   &-&  \cos(k_l+k_m-k_n)-\cos(k_n) \big] \Big\} .
\label{M-kappa}
\end{eqnarray}
Here ${\cal J}_0$ is the zeroth Bessel function, and  
$k_u = 2 \pi u /N$ denote the lattice momenta. Unlike the case of standard
potential driving, in which the renormalized Hamiltonian in the 
high-frequency limit is described by just a single
parameter $\Jeff$, the Hamiltonian \eqref{eq:ham_eff} involves a variety of matrix elements, their number and value depending on the size of the lattice. 
%[see Eq. (\ref{Q-wxyz})].

From Eqs. (\ref{eq:ham_eff})-(\ref{M-kappa}) it is clear that the only remaining energy scale in the
problem is $U$. Thus changing $U$ in the original Hamiltonian
(\ref{eq:driven_hubb_ham}) simply results in a rescaling of time, without
introducing different behaviour. We have verified this by explicit simulation
of the full time-dependent Hamiltonian (\ref{eq:driven_hubb_ham}).
We have also compared the energy spectrum of $\Heff$ with the quasienergies
obtained from the full time-dependent model 
for several lattice sizes and values of $\kappa$, 
and find excellent agreement between the two
\footnote{Comparing the quasienergies $\epsilon_j$ and the eigenenergies
of the effective Hamiltonian $E^\mathrm{eff}_j$ by evaluating 
$\chi^2 = \sum |\epsilon_j - E^\mathrm{eff}_j |^2$,
we find that for a fixed value of $\kappa$ it is bound by
$\chi^2 < 5\times 10^{-7}$. The ordering and the values of the quasienergies
are thus essentially indistinguishable from the energy levels of the
effective Hamiltonian.}.

\begin{figure}
\begin{center}
\includegraphics[width=0.43\textwidth,clip=true]{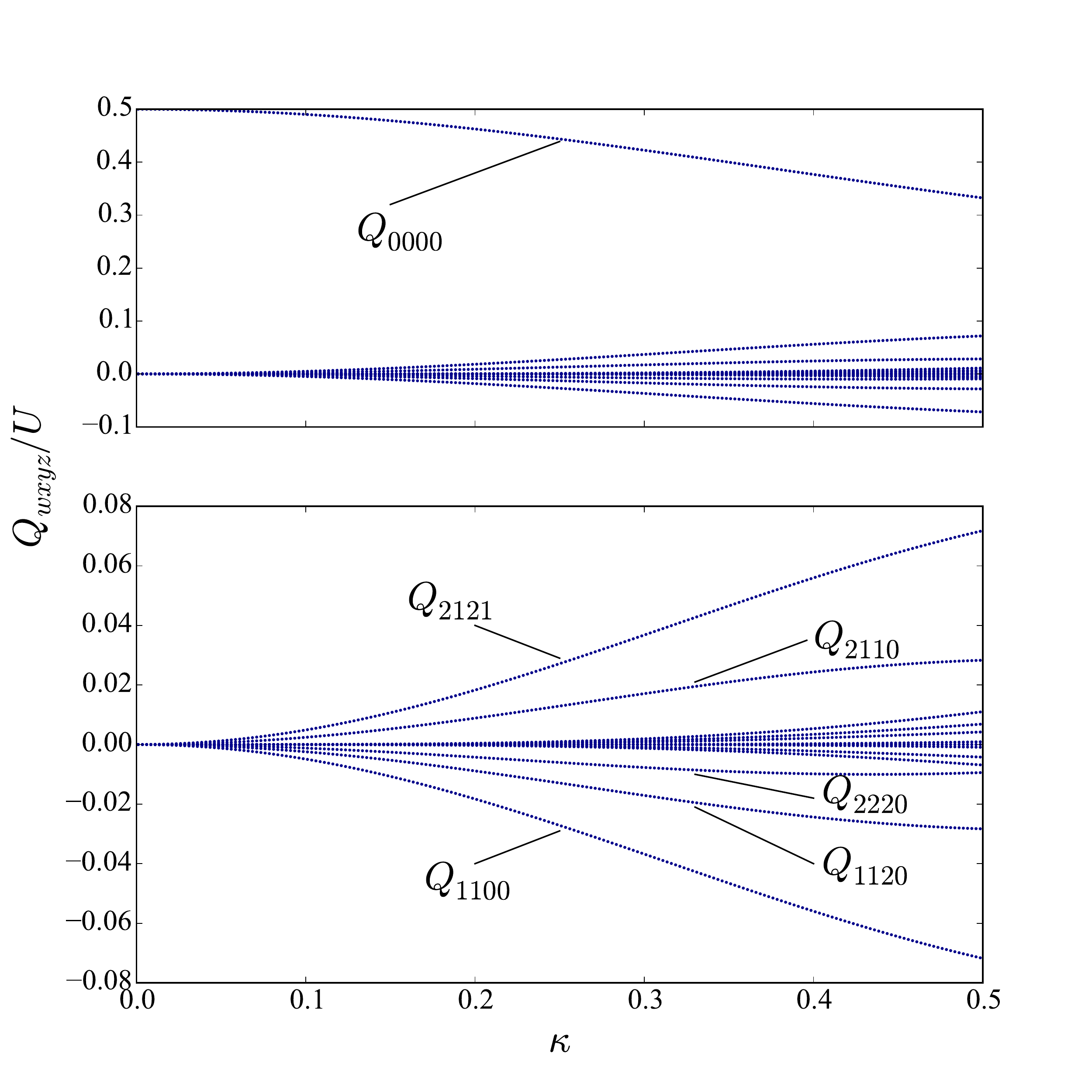}
\includegraphics[width=0.35\textwidth,clip=true]{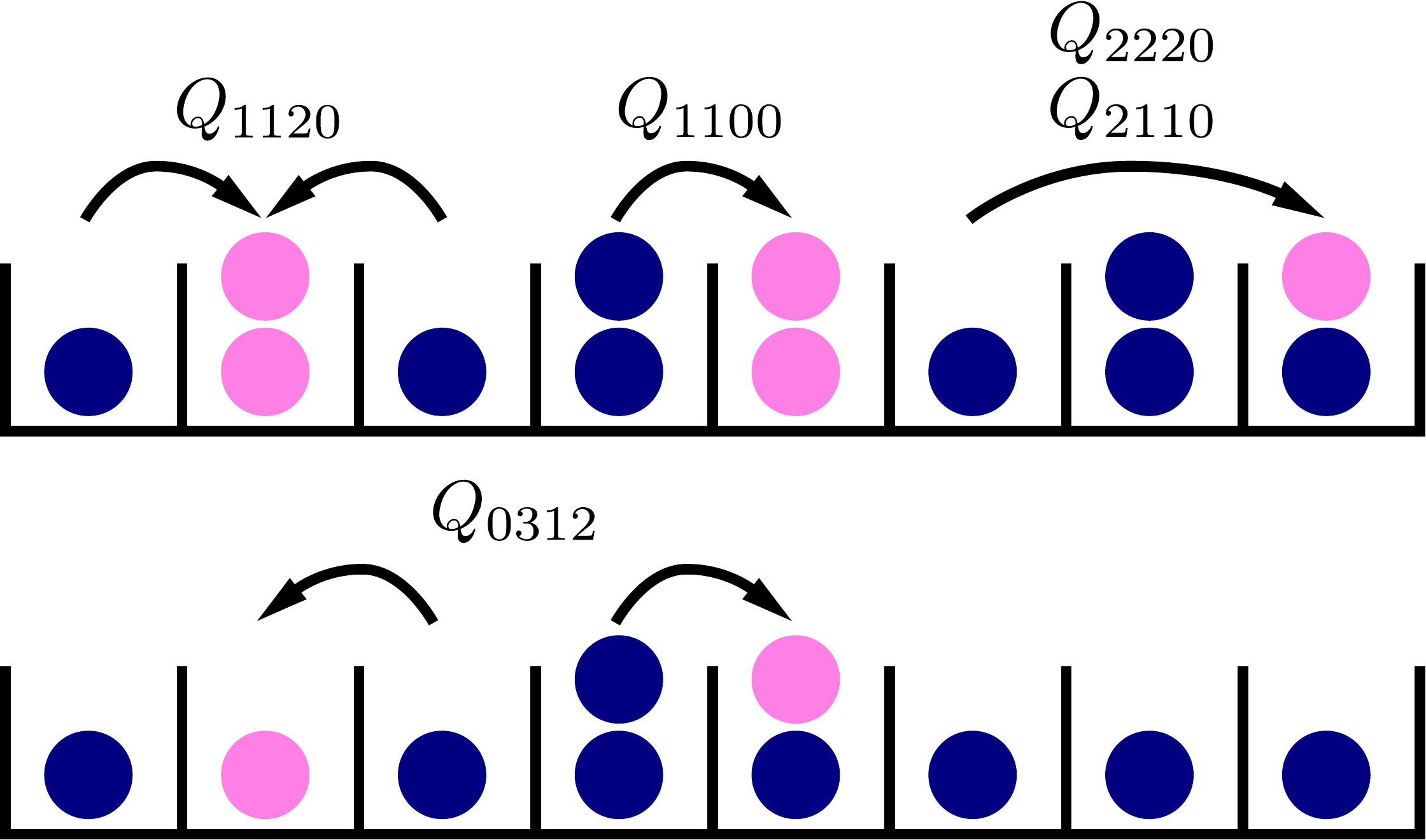}
\end{center}
\caption{{\it Top}: All hopping/interaction energies for an 8-site model in units of $U$.
The dominant term is
$Q_{0000}$, which is the renormalized on-site interaction energy. {\it Middle}: Magnification of the non-dominant processes; $Q_{wxyz}$ represent a 
given class of amplitudes. $Q_{2121}$ is the nearest neighbor interaction energy, 
the other amplitudes are illustrated below.
{\it Bottom}: Magenta (light gray) particles indicate the new position after a hopping event.}
\label{effectiveq}
\end{figure}

We can also see that although the time-averaging removes all nearest-neighbor 
single-particle hopping, higher-order processes are permitted in $\Heff$, 
and can be of unusual type, with a variety of elementary processes that blur the distinction between interaction, correlated hopping, and assisted hopping. For example, while
$Q_{0000}$ and $Q_{2121}$ are standard density-density interaction
terms, $Q_{1100}$ represents hopping of a doublon, $Q_{1120}$ provides another instance of correlated hopping, while
$Q_{wxxy}$ represents assisted hopping in which the jump between
sites $w$ and $y$ is influenced by the occupation of site $x$. 
It is important to note that hopping can in principle be arbitrarily long-ranged,
as has also been seen in studies of the driven-tunneling Kitaev model \cite{monica}.
However, for small $\kappa$ only a few hopping processes are significant,
as shown in Fig. \ref{effectiveq}, which makes analyzing the system
tractable in this range.

\section{\label{sec_results} Results}

\subsection{Momentum density function}
\begin{figure}
\begin{center}
\includegraphics[width=0.40\textwidth,clip=true]{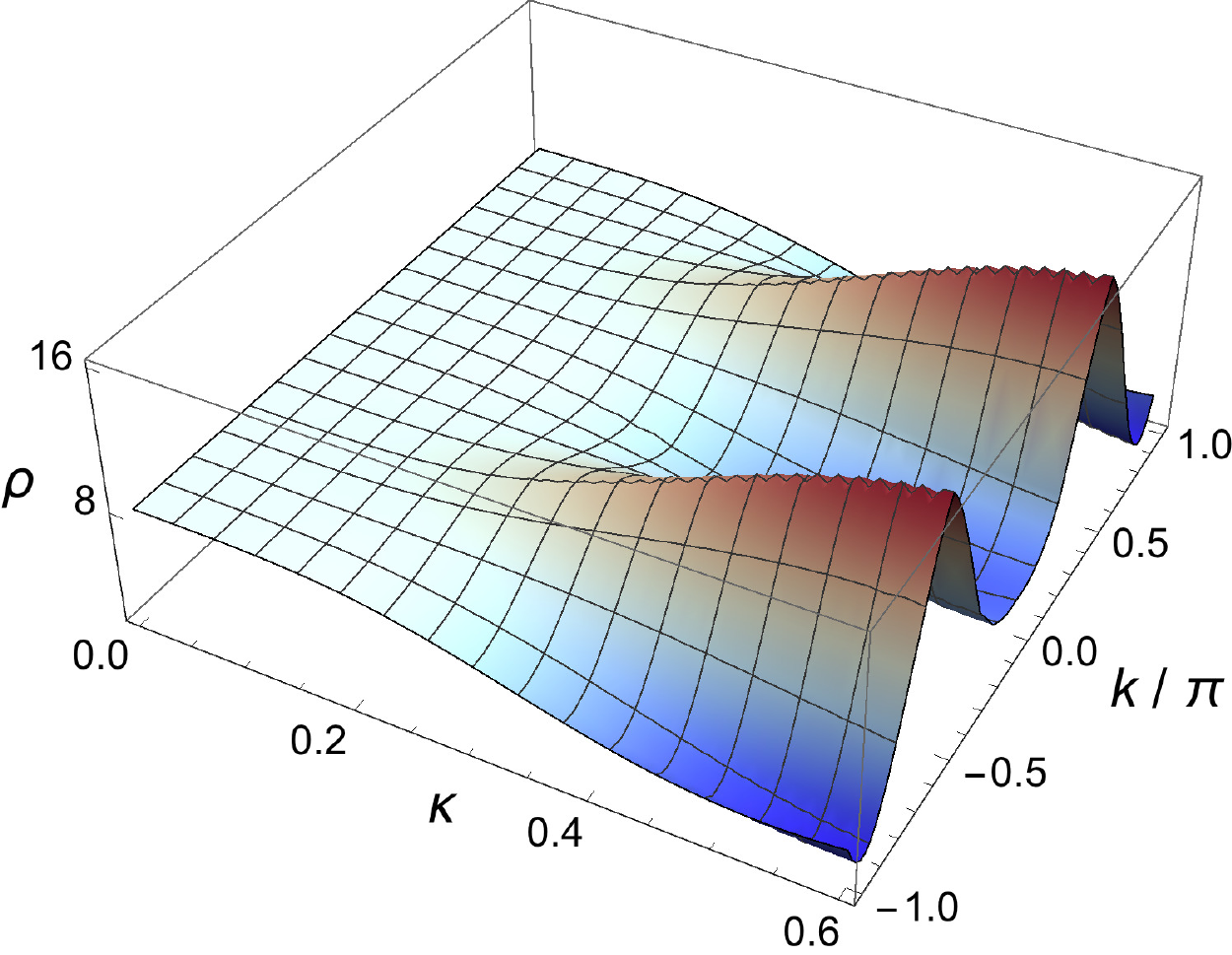}
\end{center}
\caption{Momentum density $\rho(k)$ for different values of $\kappa$,
for a system of 8 particles on 8 sites.
When $\kappa=0$, the system is in the Mott state
and $\rho(k)$ is a constant. As $\kappa$ is increased,
the momentum density function develops peaks
at $k = \pm \pi/2$, indicating the formation of
condensate states with non-zero momentum. In this plot we smoothly interpolate between the 8 allowed
values of lattice momentum.
}
\label{mom_den}
\end{figure}

\begin{figure}
\begin{center}
\includegraphics[width=0.45\textwidth,clip=true]{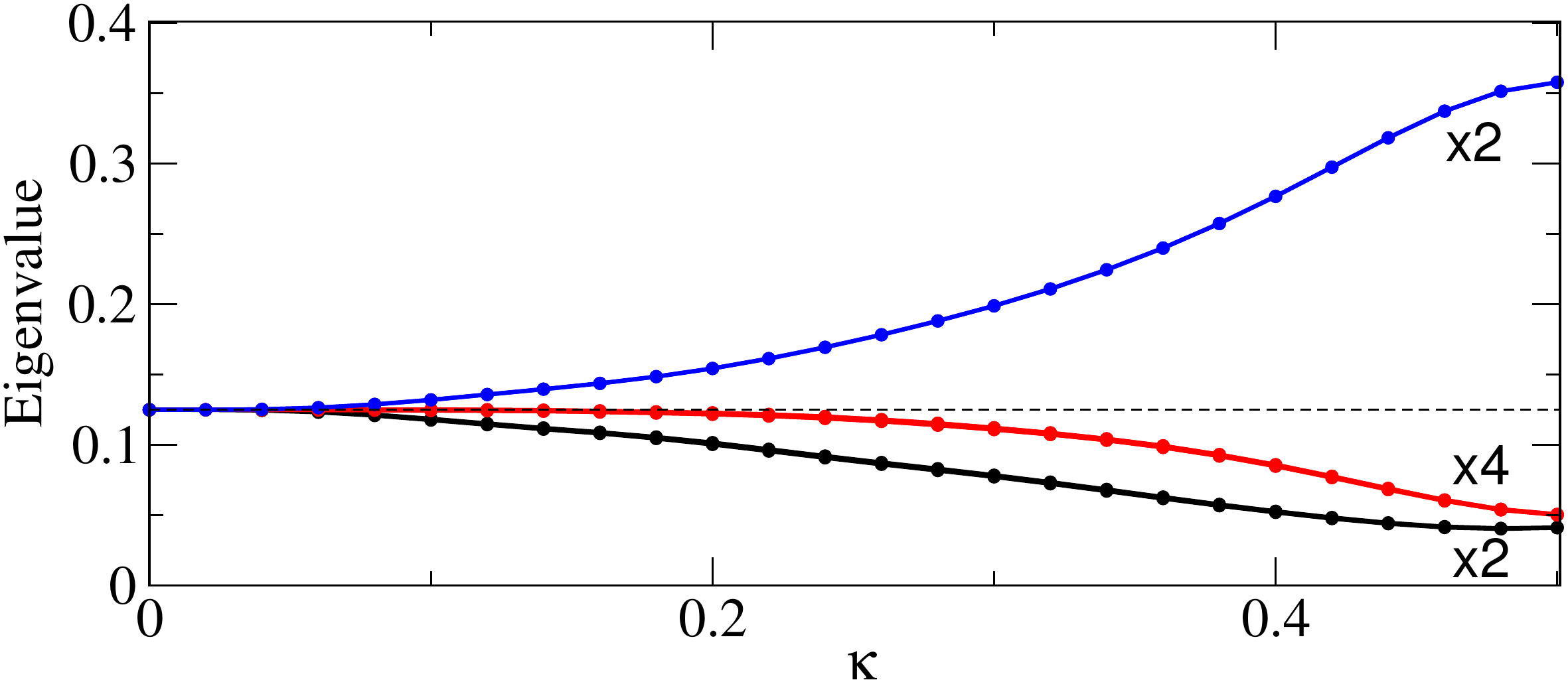}
\end{center}
\caption{Eigenvalues of the one-particle reduced density matrix.
At $\kappa = 0$ all the eigenvalues are equal, implying
the even spread of the Mott state over all orbitals.
As $\kappa$ increases, two (degenerate) eigenvalues increase at
the expense of the others,
signifying the macroscopic occupation of these two natural orbitals, and
thus the formation of a fragmented condensate. The degeneracy of
each set of eigenvalues is labeled in the figure.}
\label{orbitals}
\end{figure}

In Fig. \ref{mom_den} we show the momentum density function
for an eight-site system,
\begin{equation}
\rho(k) = \sum_{j, l} e^{\im k \left( r_j - r_l \right)}
P^{(1)}_{j l} ~,
\end{equation}
where $P^{(1)}_{j l} = \langle a_j^\dagger a_l \rangle$ is 
the one-particle reduced density matrix. 
For $\kappa = 0$ the system is in the Mott state and $\rho(k)$ is
completely flat, reflecting the absence of phase coherence. As $\kappa$
is increased, however, peaks develop at $k = \pm \pi/2$, suggesting the
formation of condensates at those momenta \footnote{For simplicity of language, here and in the following we employ the term ``condensate'' although, strictly speaking, in one-dimension only a ``quasi-condensate'' exists.}. 
To further explore the nature of these peaks, we calculate the eigensystem of
the one-particle reduced density matrix.
Its eigenstates are known as natural orbitals, and the occupation of each
natural orbital is given by its corresponding eigenvalue, a condensate being 
formed if an orbital is macroscopically
occupied. In Fig. \ref{orbitals} we can see that the natural orbitals are equally
occupied for $\kappa = 0$, just as expected for the Mott state.
Comparing Fig. \ref{mom_den} with Fig. \ref{orbitals}, the formation of
the peaks in the momentum density is clearly correlated with the enhanced occupation
of two (degenerate) natural orbitals, indicating the formation of a fragmented
condensate. Numerical inspection shows that this pair of degenerate orbitals very much resemble plane waves of momentum $\pm \pi/2$. 
This contrasts with the behaviour of the standard Mott-superfluid transition 
(described by the undriven BH model), for which a single condensate develops 
at $k=0$. 

Examining the ground state of $\Heff$ over the displayed range of $\kappa$ reveals
that it is strongly dominated by the Mott state, with the next most important
contributions coming from ``dipole states''. These are states containing
a doublon-hole pair separated by a single site
\begin{equation}
|\Psi_{\rm dip}\rangle = \frac{1}{\sqrt{2 N}}\sum^{N-1}_{x=0} \left(D_{x}+D^\dagger_{x}\right) 
|{\rm Mott}\rangle~,
\label{dipole}
\end{equation}
where $D_x = a^\dagger_{x}a_{x+2}$.
Inspection of the numerical results suggests approximating the
ground state by
\begin{equation}
|\xi\rangle = \alpha |{\rm Mott}\rangle +\beta |\Psi_{\rm dip}\rangle~,
\label{eq:ansatz}
\end{equation}
where $\alpha,\beta \in \mathbb{R}~$ and $|\alpha| > |\beta|$ is expected.
Although inspired by the numerical simulation of a finite-size system, we emphasize that the state (\ref{eq:ansatz}), as well as the following discussion, 
has a well-defined thermodynamic limit. 

The expectation value of (\ref{eq:ham_eff}) in the state (\ref{eq:ansatz}) is 
\begin{eqnarray}
\langle H_{\rm eff}\rangle =& \alpha^2 \langle{\rm Mott}|H_{\rm eff}|{\rm Mott}\rangle
+\beta^2 \langle \Psi_{\rm dip} |H_{\rm eff} |\Psi_{\rm dip}\rangle  \nonumber \\
&+ 2\alpha\beta \ {\rm Re}\, \langle {\rm Mott}| H_{\rm eff} |\Psi_{\rm dip}\rangle  
\label{eq:eff_ham_dip_exp}
\end{eqnarray}
which suggests that if ${\rm Re\,}\langle {\rm Mott}| H_{\rm eff} |\Psi_{\rm dip}\rangle > 0$,
the energy is minimized for $\mathrm{sgn}(\alpha \beta) = -1$. As the Mott and dipole states are only connected by matrix elements of the type $Q_{2xx0}$, 
we obtain
\begin{align}
\begin{aligned}
&\langle {\rm Mott}| H_{\rm eff} |\Psi_{\rm dip}\rangle =2N\sqrt{2} 
\Big( \sum_{x\neq0}+\sum_{x\neq 0, \, 2}\Big)Q_{2xx0} \ ,
\label{eq:mott_dip_me_eff}
\end{aligned}
\end{align}
and we have verified that this is indeed positive (see middle Fig. \ref{effectiveq})
over the range of $\kappa$ that we investigate, for chains of up to 90 sites. 
The negative sign of $\left( \alpha \beta \right)$ also
causes the momentum density 
\begin{equation}
\langle \xi | \rho(k) |\xi\rangle 
 = N \left( 1+\frac{4 \alpha \beta}{\sqrt{N}} \cos (2 k) + 
\frac{3\beta^2}{N}\cos (4 k) \right)~ 
\label{eq:mom_dens_ansatz}
\end{equation}
to develop peaks at $k=\pm \pi/2$,
in agreement with the numerical results shown in Fig. \ref{mom_den}.
  
The presence of density peaks at nonzero momenta can be further understood physically
by noting that the predominantly positive character of the matrix elements.
$Q_{yxx,y+2}$ corresponds, in the usual convention of Eq. (\ref{eq:driven_hubb_ham}), to second-order hopping with a {\em negative} effective mass. 
This favors the occupation of single-particle states with a phase difference of $\pi$ between next-nearest neighbors, which are states with momentum $\pm \pi/2$.
For higher driving strengths, however, we find that this simple picture of Eq. \eqref{eq:ansatz} breaks down due to the increasing weight of other configurations in the ground state.

Figure \ref{fig:dip_ener_comp} in Appendix \ref{app_dip} shows the ground-state energy as a function of the driving parameter $\kappa$. We observed that, comparing with the exact calculation, the dipole-approximation gives good results up to $\kappa \simeq 0.3$. For that value of $\kappa$, Fig. \ref{mom_den} shows that the momentum distribution is beginning to peak at momenta $k=\pm \pi/2$. 
Later in this section we will see that the insulator-superfluid transition occurs at $\kappa_c \simeq 0.42$. Thus, although the dipole approximation accounts for a partial condensation of bosons at momenta  $\pm \pi/2$ [see Eq. (\ref{eq:mom_dens_ansatz})], its range of validity remains confined to the insulator region.

\subsection{Two-particle momentum density}
\label{2-particle-momentum}

\begin{figure}
\begin{center}
\includegraphics[width=0.35\textwidth,clip=true]{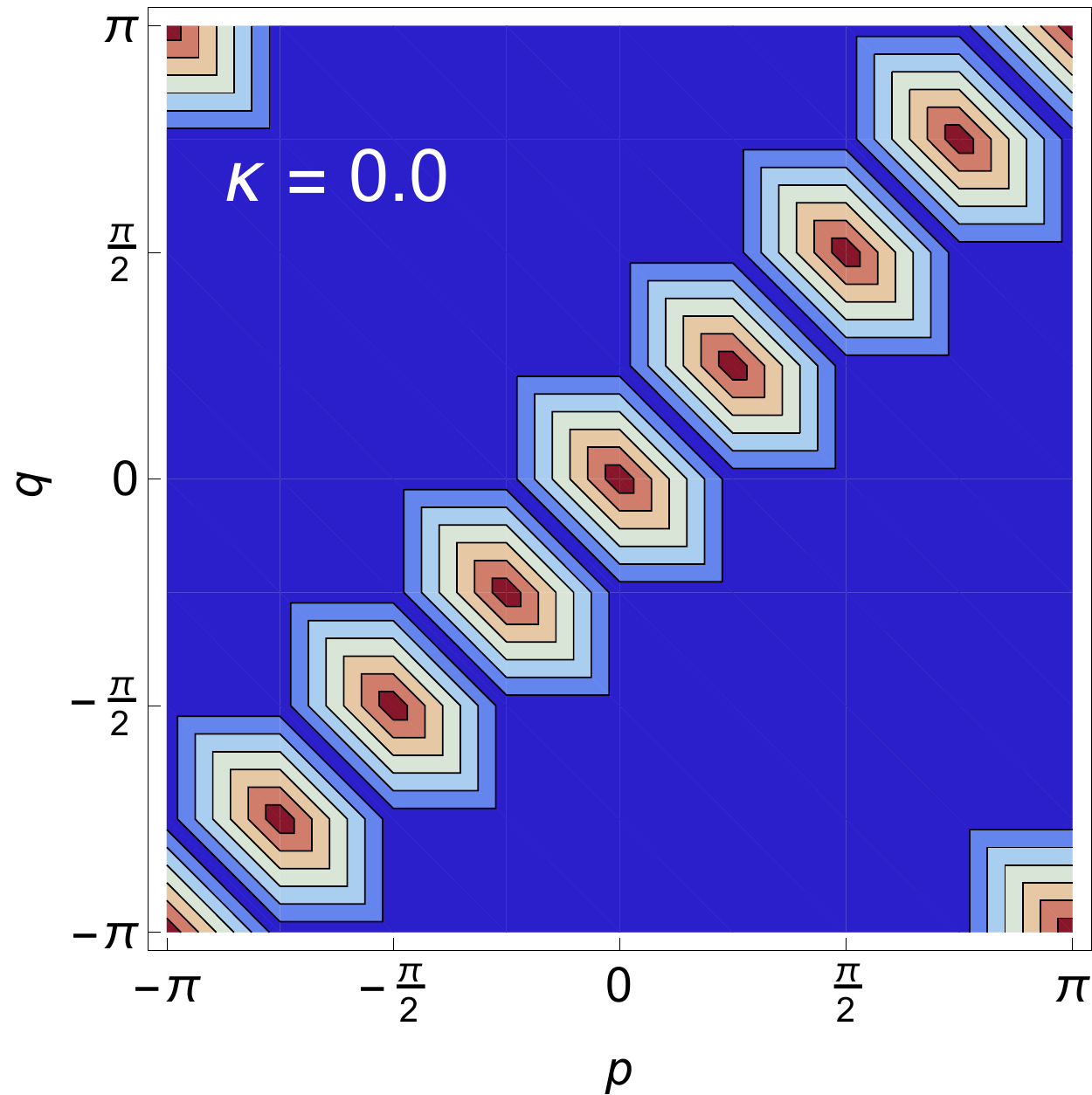}
\includegraphics[width=0.35\textwidth,clip=true]{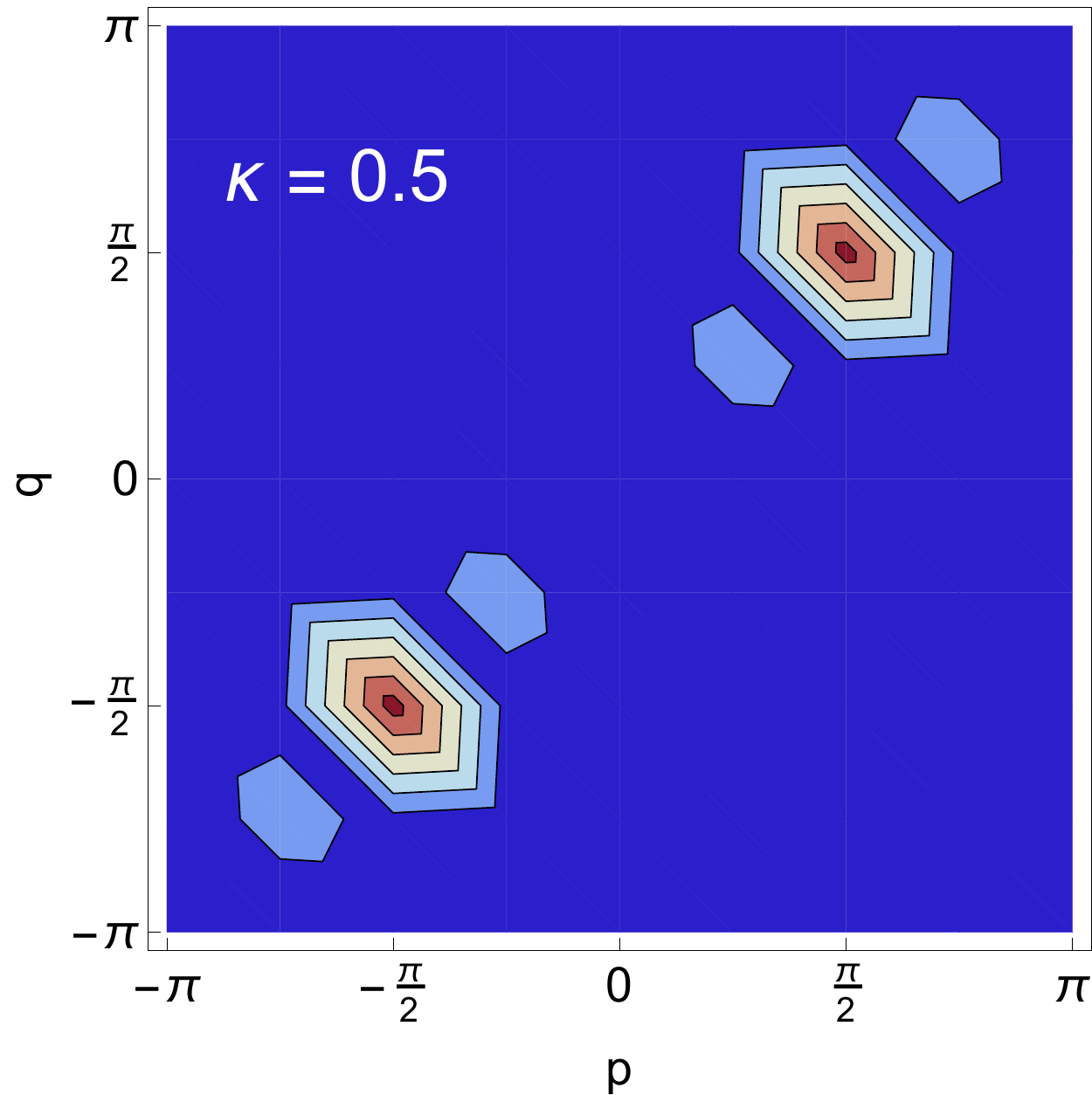}
\end{center}
\caption{Two particle reduced density matrix,
$\rho^{(2)}(p,q) = \langle \rho(p) \rho(q) \rangle$, for two values of the
driving parameter $\kappa$ for an 8-site system. Above: $\kappa = 0$. The
system is a perfect Mott insulator, and $\rho^{(2)}(p,q)$
consists of a uniform background, with a line of peaks along
$p = q$. Below: $\kappa = 0.5$.
$\rho^{(2)}(p,q)$ now shows two peaks, centered on
$(-\pi/2, -\pi/2)$ and $(\pi / 2, \pi / 2 )$.
This corresponds to the system being in a cat state; a
superposition of $N$ particles peaked at momentum $-\pi/2$, and
$N$ particles peaked at $\pi/2$.}
\label{2particle_RDM}
\end{figure}

The behaviour of the one-particle reduced density matrix clearly
indicates that the system makes a transition to a fragmented condensate.
However, to characterize this state more precisely,
higher-order correlation functions are required \cite{mueller}, in particular
the two-particle reduced density-matrix,
$P^{(2)}_{j k l m} = \langle a_j^\dagger a_k a_l^\dagger a_m \rangle$.
In Fig. \ref{2particle_RDM} we show the Fourier transform of this
quantity
\begin{equation}
\langle \rho(p) \rho(q) \rangle = \sum_{j,k,l,m}
e^{\im p \left( r_j - r_k \right)}
e^{\im q \left( r_l - r_m \right)} P^{(2)}_{j k l m} \ ,
\label{FT_2RDM}
\end{equation}
for two different values of the driving strength $\kappa$.
For $\kappa = 0$ the system is a perfect Mott insulator,
and we can see from Fig. \ref{2particle_RDM}a
that $\langle \rho(p) \rho(q) \rangle$ manifests peaks along
the line $p = q$, and is flat elsewhere. For the case of a perfect
Mott state, $\langle \rho(p) \rho(q) \rangle$ can be calculated
analytically, yielding the result
\begin{equation}
\langle \rho(p) \rho(q) \rangle = N \left( N - 2 \right)
+ N^2 \ \delta_{p q} \ ,
\label{analytic_RDM}
\end{equation}
which for momenta discretised over
the first Brillouin zone gives
perfect agreement with the numerical results.

Below, in Fig. \ref{2particle_RDM}b, we show the result for $\kappa =0.5$,
for which the system is well within the regime in which the condensate develops.
Unlike the Mott state, only two peaks are visible, located at
$(p,q) = (\pi/2, \pi/2)$ and $(p,q) = (-\pi/2, -\pi/2)$.
This again highlights the fragmented nature of this condensate; in the
standard case of the Mott transition, the superfluid
would present just a single peak
in $\langle \rho(p) \rho(q) \rangle$, centered on the origin.

In addition to confirming the fragmentation of the condensate,
Fig. \ref{2particle_RDM}b provides information
about the nature of the fragmentation. 
A Fock state in momentum space, having
the approximate form
\begin{equation}
| \mathrm{Fock} \rangle =
\frac{1}{(N/2)!}
\left( a_{\pi/2}^\dagger \right)^{N / 2} \left( a_{-\pi/2}^\dagger \right)^{N / 2}
| 0 \rangle \, ,
\label{eq:Fock-state}
\end{equation}
would present four peaks in the two-particle momentum density,
located at the points $( \pm \pi/2, \pm \pi/2 )$. 
However, as Fig. \ref{2particle_RDM}b shows two peaks the ground state
is clearly {\em not} of this type. The absence of the anti-diagonal peaks
in the two-particle momentum density
instead suggests that the ground state has a Schr\"odinger-cat 
structure, which in its idealized form would be
\begin{equation}
| \mathrm{cat} \rangle =
\frac{1}{\sqrt{2N!}}
\left[(a_{\pi/2}^\dagger)^N + (a_{-\pi/2}^\dagger)^N \right]| 0 \rangle \, .
\label{eq:cat-state}
\end{equation}
This finding has considerable implications for the observation
of this state in experiment \cite{marco}. While a single measurement of the 
momentum distribution of a Fock state would yield an equal weight for the 
two momenta $k = \pm \pi/2$, the cat state would be collapsed to one 
momentum or the other with equal probability. 
This unusual behaviour would thus provide a clear experimental
signal of the exotic superfluid state that this form of driving
will produce.

\subsection{Vanishing of the Mott gap}
\label{Mott-gap-subsection}
\begin{figure}
\begin{center}
\includegraphics[width=0.45\textwidth,clip=true]{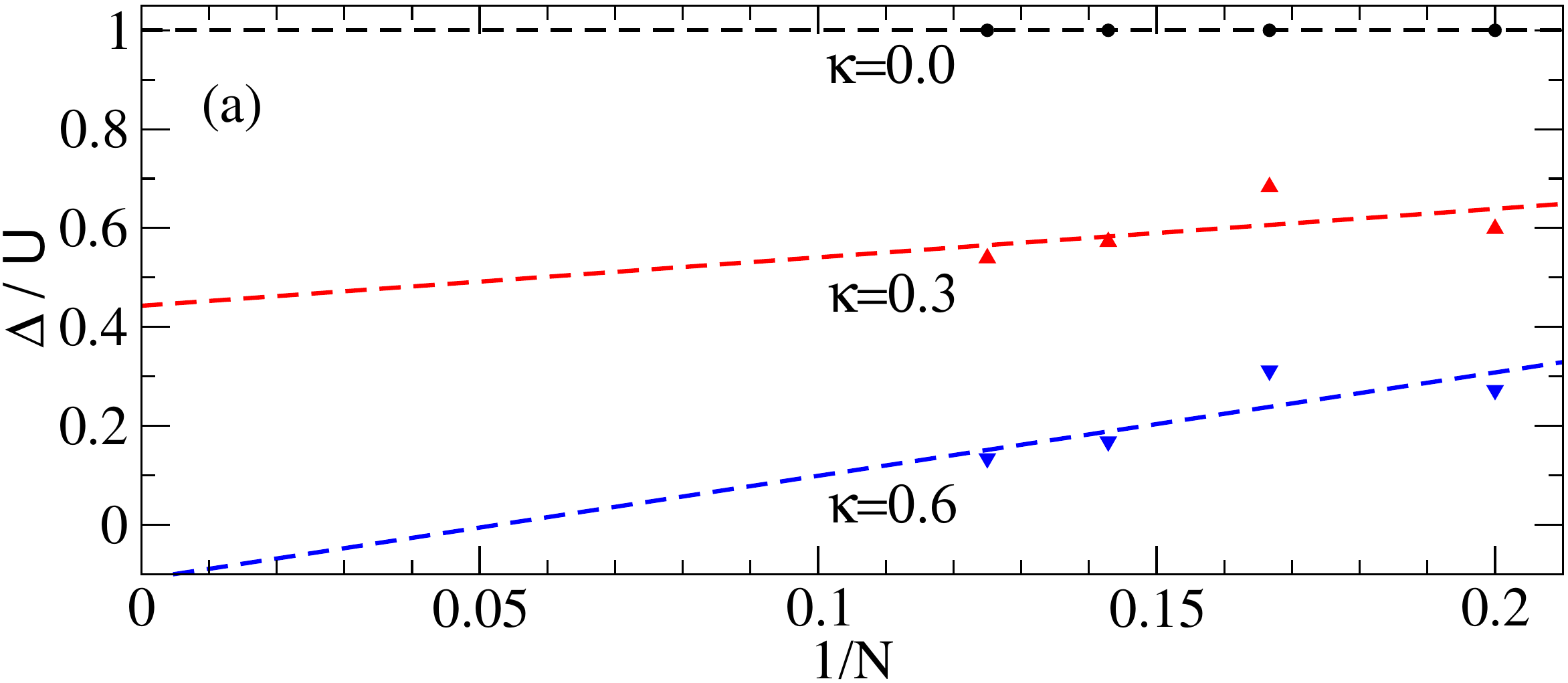}
\includegraphics[width=0.45\textwidth,clip=true]{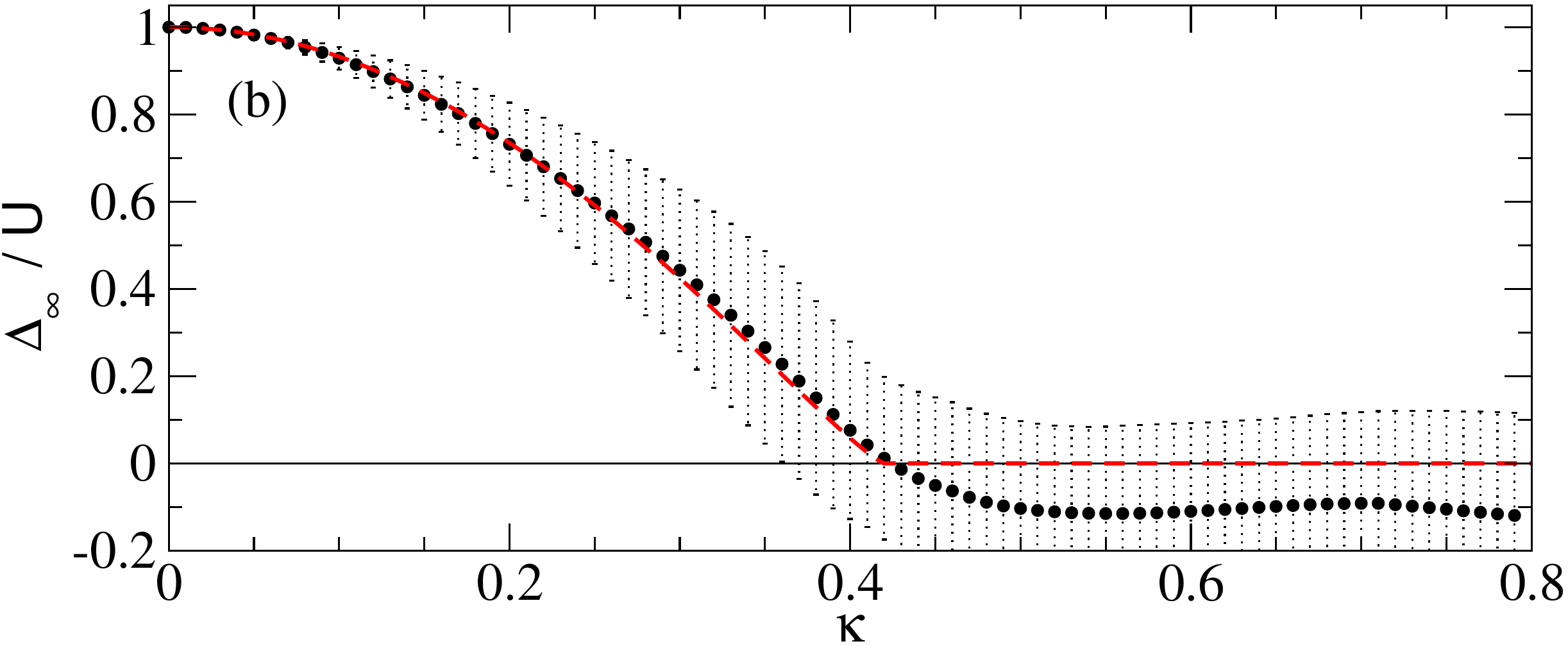}
\includegraphics[width=0.45\textwidth,clip=true]{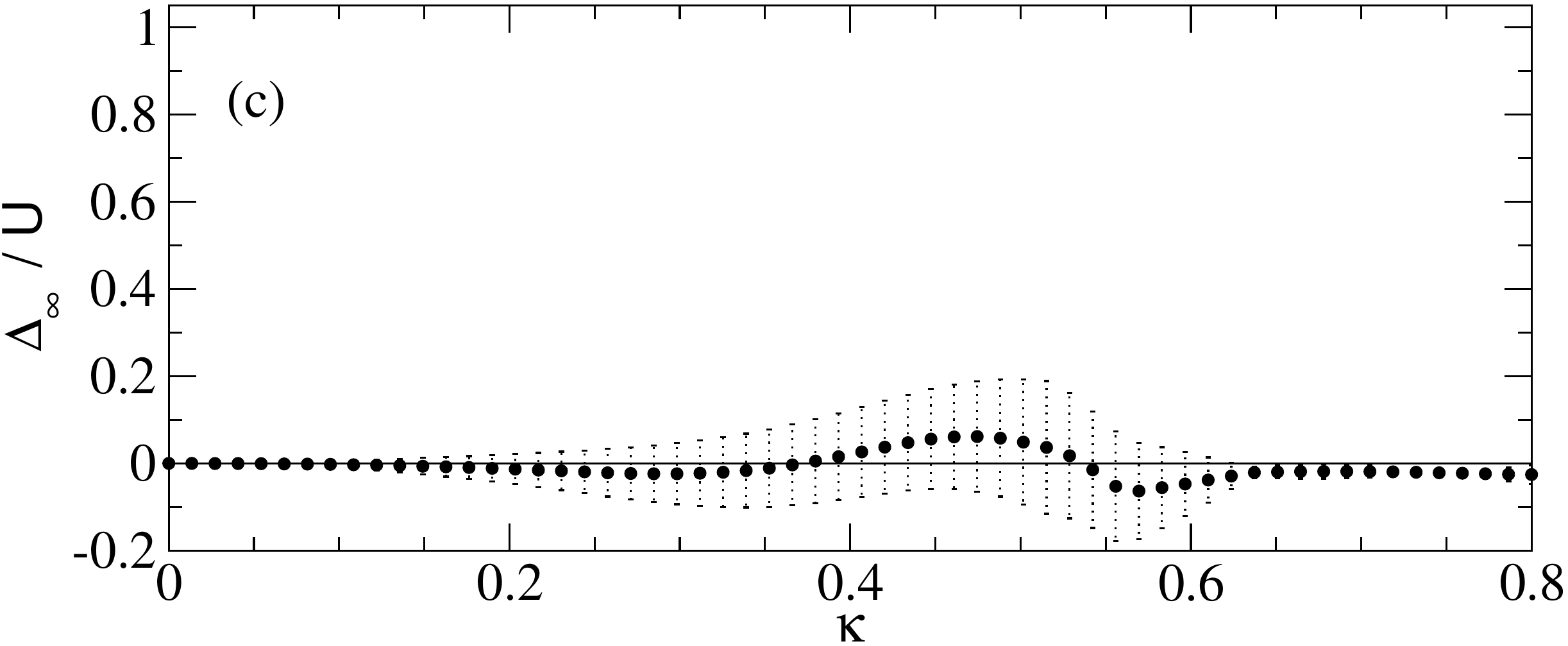}
\end{center}
\caption{(a) Behaviour of the gap, $\Delta(N)$ as a function of the
lattice size $1/N$. In each case the number of bosons is
equal to the number of lattice sites (commensurate filling).
The numerical uncertainty in the data-points is smaller
than the symbol size.
(b) The value of $\Delta$ in the thermodynamic limit, $\Delta_{\infty}$,
is obtained from a least-square fit to the data shown above; the
error bars indicate the statistical uncertainty in the fitting procedure.
At $\kappa = 0.42$
the gap vanishes, indicating a phase transition from the Mott phase
to a superfluid state. The (red) dashed line is a guide to the eye, following
the power law
$\Delta / U = \left( 1 - ( \kappa / \kappa_c)^2 \right)^{1.2}$.
(c) Behaviour of the gap, $\Delta_{\infty}$ obtained in the same way
as in (b),
but for incommensurate filling. In contrast to the commensurate case, the 
system stays gapless, i.e. superfluid, over the entire range of driving. 
}
\label{fig_gap}
\end{figure}

The Mott phase is characterized by a non-zero energy gap, $\Delta$, for adding
or subtracting a particle from the system. Conversely, a superfluid state 
will be gapless. As a result, evaluating $\Delta$ and locating
the point at which $\Delta = 0$, provides a direct method of identifying the
point at which the phase transition occurs.
To calculate $\Delta$ we follow the procedure described in Ref. \cite{sowinski},
where it was applied to the standard Bose-Hubbard model. We first
obtain the ground-state energy for the case of commensurate filling,
when the number of lattice sites, $N$, is equal to the number of bosons.
We then define the energy gap to be the difference
\begin{equation}
\Delta(N) = E(N,N+1) - 2 E(N,N) + E(N,N-1) \ ,
\label{eq_gap}
\end{equation}
where $E(m,n)$ denotes the ground-state energy of an $m$-site system
holding $n$ bosons. In order to evaluate $\Delta$ in
the thermodynamic limit, we diagonalise the many-body effective Hamiltonian
\eqref{eq:ham_eff} for several different lattice sizes ($N = 5,\dots,8$),
and make a least-square fit to the function
\begin{equation}
\Delta(N) = \Delta_{\infty} + d / N + {\cal O}(1/N^2) \ ,
\end{equation}
to extract the value of $\Delta$ in the $1/N = 0$ limit.
Although these systems may appear rather small, Ref. \cite{sowinski}
showed that this technique indeed provides a robust and well-controlled
means of extrapolating to the thermodynamic limit, even for
systems of this size.
In Fig. \ref{fig_gap}a we plot $\Delta(N)$ as a function of $1/N$ 
for several values of $\kappa$, and indeed see an approximately linear 
behaviour. We show in Fig. \ref{fig_gap}b the behaviour of $\Delta_{\infty}$,
obtained in this way, as $\kappa$ is varied.
Although the deviations from linearity in the fitting procedure
introduce rather large error bars, the 
tendency of $\Delta_{\infty}$ is clear. 
In the absence of driving ($\kappa = 0$), $\Delta_{\infty} = U$ as 
expected. As $\kappa$ is increased, $\Delta_{\infty}$ reduces,
and reaches zero at approximately $\kappa = 0.42$. Thereafter
$\Delta_{\infty}$ remains zero within the error bars, indicating that the Mott gap has
closed, and that the system is in the superfluid regime. 

As a further test of the method, we consider the case of an 
{\em incommensurate} filling, focusing on the case of a system of $N$ sites with
$N-2$ particles, for which the gap is given by
\begin{equation}
\Delta^{\rm ic}(N) = E(N,N-1) - 2 E(N,N-2) + E(N,N-3) \, .
\label{eq_gap-incomm}
\end{equation}
%($N_\mathrm{boson} / N = 3 / 4$).
Analyzing this 
system in the same way (computing the cases $N=5,\dots,8$) yields a value for the gap that remains zero 
(within the error bars) for all driving strengths, indicating that the system 
is {\em always} superfluid. This finding anticipates a result of Luttinger liquid
theory \cite{cazalilla}, which states that for non-commensurate fillings a 
one-dimensional system falls into a different universality class than for the
commensurate case,
and will be superfluid for all non-zero values of $\kappa$.
This property is also evident in the behaviour of the Luttinger
parameter $K_b$ (shown in Fig. \ref{lutt}b), as we will discuss in the
next section.

\subsection{Luttinger liquid parameters}
The unifying concept for one-dimensional interacting systems, of both bosonic
and fermionic type, is the Tomonaga-Luttinger liquid \cite{tomonaga,luttinger}. 
A system of this type exhibits two important features, namely, that
correlation functions follow a power-law decay \cite{haldane}, 
and that the low-energy excitations are collective modes with a linear dispersion.
A Tomonaga-Luttinger liquid can thus be exactly specified by just two parameters;
the interaction parameter $K_b$, in which the asymptotic behaviour of all the
correlation functions can be expressed, and the
group velocity, $v$, of the collective modes.

\subsubsection{Correlation function}
For the system we study, $K_b$ can be extracted from the
the density-density correlation function \cite{white}
\begin{equation}
N(r) = \frac{1}{N} \sum_{x=1}^N \langle n_x n_{x+r} \rangle
-\langle n_x \rangle \langle n_{x+r} \rangle \ ,
\end{equation}
which asymptotically decays as \cite{cazalilla}
\begin{equation}
N(r \rightarrow \infty) = \frac{-1}{2 K_b \left( \pi r \right)^2} +
A \frac{\cos \left(2 \pi r \right)}{ r^{2 / K_b}} + \dots \ .
\label{decay}
\end{equation}
Rather than directly fitting $N(r)$ to Eq. (\ref{decay}), however, more stable results
are obtained by evaluating its Fourier transform, $N(k)$, and estimating
$K_b$ from its derivative \cite{gebhard}
\begin{equation}
\frac{1}{2 \pi K_b} = \mathrm{lim}_{k \rightarrow 0} \frac{N(k)}{k}~.
\label{Kb}
\end{equation}

\begin{figure}
\begin{center}
\includegraphics[width=0.45\textwidth,clip=true]{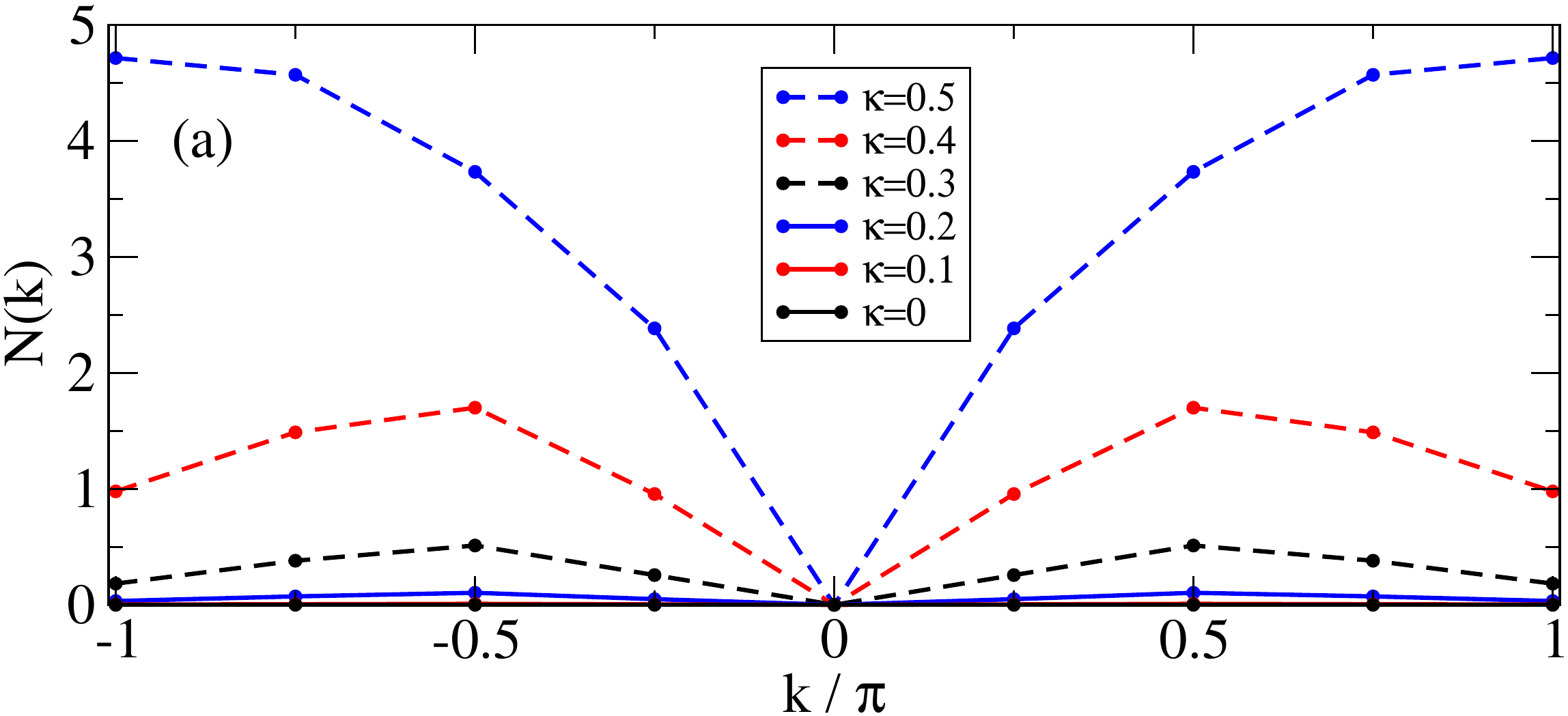}
\includegraphics[width=0.45\textwidth,clip=true]{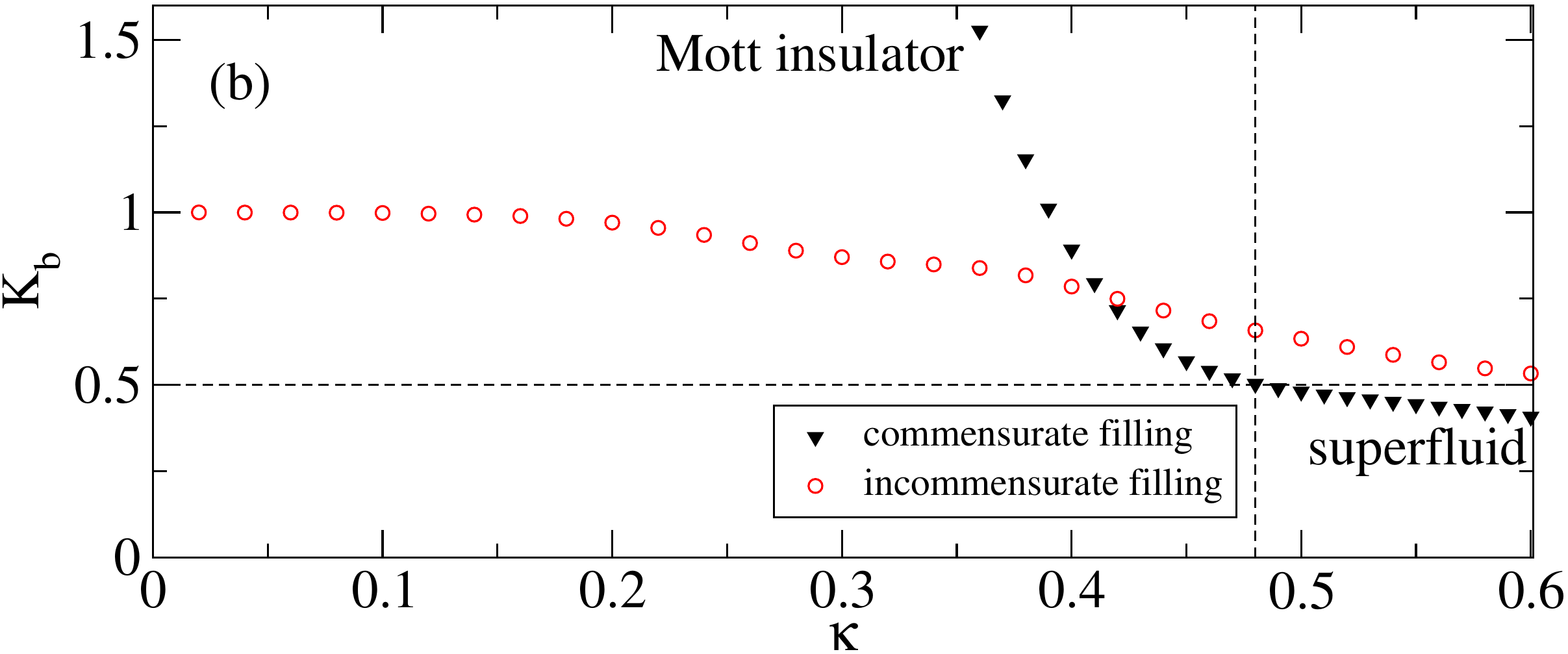}
\end{center}
\caption{a) Density correlation function in momentum
space for various driving amplitudes $\kappa$ in the commensurate case of 8 particles 
in 8 sites. For small values of $\kappa$ (solid lines) the system is gapped,
and so $N(k)$ is suppressed. For larger $\kappa$ (dashed lines) the
system is gapless, and $N(k)$ has a linear behaviour for small $k$. 
b) Luttinger parameter, $K_b$, calculated
from the density correlation function (see Eq. (\ref{Kb})).
For commensurate filling (black triangles), $\kappa$, $K_b$ diverges as $\kappa$ reduces, indicating a Mott insulator.
When $K_b$ passes through the universal value $K_b = 1/2$, the system makes
a transition to a superfluid state. By contrast, an incommensurate system of 6 bosons in an 8-site
lattice (red circles) is superfluid for all non-zero values of $\kappa$. In this case, as $\kappa \rightarrow 0$,
$K_b$ approaches unity, as expected for a system in this universality class.}
\label{lutt}

\end{figure}

In Fig. \ref{lutt}a we show $N(k)$ for a range of values of $\kappa$.
For large $\kappa$, for which the system is gapless,
the results show a clearly linear behaviour for small $k$, indicating
that this procedure is valid. As $\kappa$ 
is reduced, however, $N(k)$ is suppressed and has a quadratic behaviour
for small $k$, corresponding to the system being gapped and thus
not being describable in terms of a Luttinger liquid.

For small values of $\kappa$ the system is in the Mott insulator regime, and
consequently $K_b$ should diverge. In Fig. \ref{lutt}b we can see that
the calculated values of $K_b$ are indeed large, but remain finite 
since the system is far from the thermodynamic limit. As $\kappa$ is increased
the value of $K_b$ drops.
For any commensurate filling, the transition between the Mott state and the
Luttinger liquid will be of Kosterlitz-Thouless type, and at the transition
$K_b$ will take a universal value.
For unit filling this is given by $K_b = 1/2$, and accordingly we can use this
as a criterion for identifying the point of the phase transition \cite{giamarchi,cazalilla}. 

We can also note that since Luttinger liquids with $K_b < 2$ are dominated
by superfluid correlations \cite{cazalilla}, 
the transition will be to a superfluid state,
exactly as we would expect from the vanishing of the gap seen previously.
From Fig. \ref{lutt}b
we can see that $K_b$ crosses the critical value at approximately $\kappa=0.48$. This
is slightly larger than the value $\kappa = 0.42$ obtained in the previous section from
the analysis of the gap, but still represents good agreement given that the
previous result was extrapolated to the thermodynamic limit, while this was
obtained on a finite system  of 8 sites.

Doping the Mott insulator away from commensurate filling has the effect of immediately
producing superfluidity. Accordingly, for an incommensurate filling the system
will only be a Mott insulator at $\kappa = 0$, and will immediately make a phase
transition as soon as $\kappa > 0$. At this point $K_b$ will again take a universal
value, in this case being exactly twice that of the commensurate transition \cite{cazalilla}. In Fig. \ref{lutt}b
we show the values of $K_b$, calculated in the same way, for an incommensurate
filling of 6 bosons on an 8-site lattice. As expected, $K_b \rightarrow 1$ in the
limit of small $\kappa$. The excellent accuracy of this limiting value of $K_b$,
obtained with no fitting parameters, provides a strong indication that this method
of extracting $K_b$ is indeed reliable. 

\subsubsection{Spectral function}

A final check of the Luttinger liquid picture is to confirm that the low-lying excitations of the
system have a linear dispersion, yielding a well-defined value for the other Luttinger parameter,
the group velocity $v$. To do this we calculate the zero temperature spectral functions for the
emission of a single particle with momentum $k$ and energy $\omega$ 
\begin{align}
	A(\omega,k) = \sum_n \left | \langle \Psi_n| a^\dagger_{k} | \Psi_g\rangle \right |^2 \delta(\omega - \omega_n)~ \, .
	\label{eq:spec_dens}
\end{align} 
Here $|\Psi_n\rangle$ is the $n$-th excited state in the $N+1$ particle system, $| \Psi_g\rangle$ is the ground state of $N$ particles and $\omega_n = E_n(N,N+1)-E_0(N,N) $, where $E_n(N,N+1)$ is the energy of the $n$-th excited state in the $N+1$ particle system. To smooth out this spectral function we convolve
Eq. \eqref{eq:spec_dens} with a narrow Lorentzian. The presence of sharp peaks in
$A(\omega,k)$ indicate the presence of well-defined quasiparticle excitations, and plotting the maxima of 
$A(\omega,k)$ allows us to determine their dispersion relation $\omega(k)$. 

\begin{figure}
\begin{center}
\includegraphics[width=.45\textwidth]{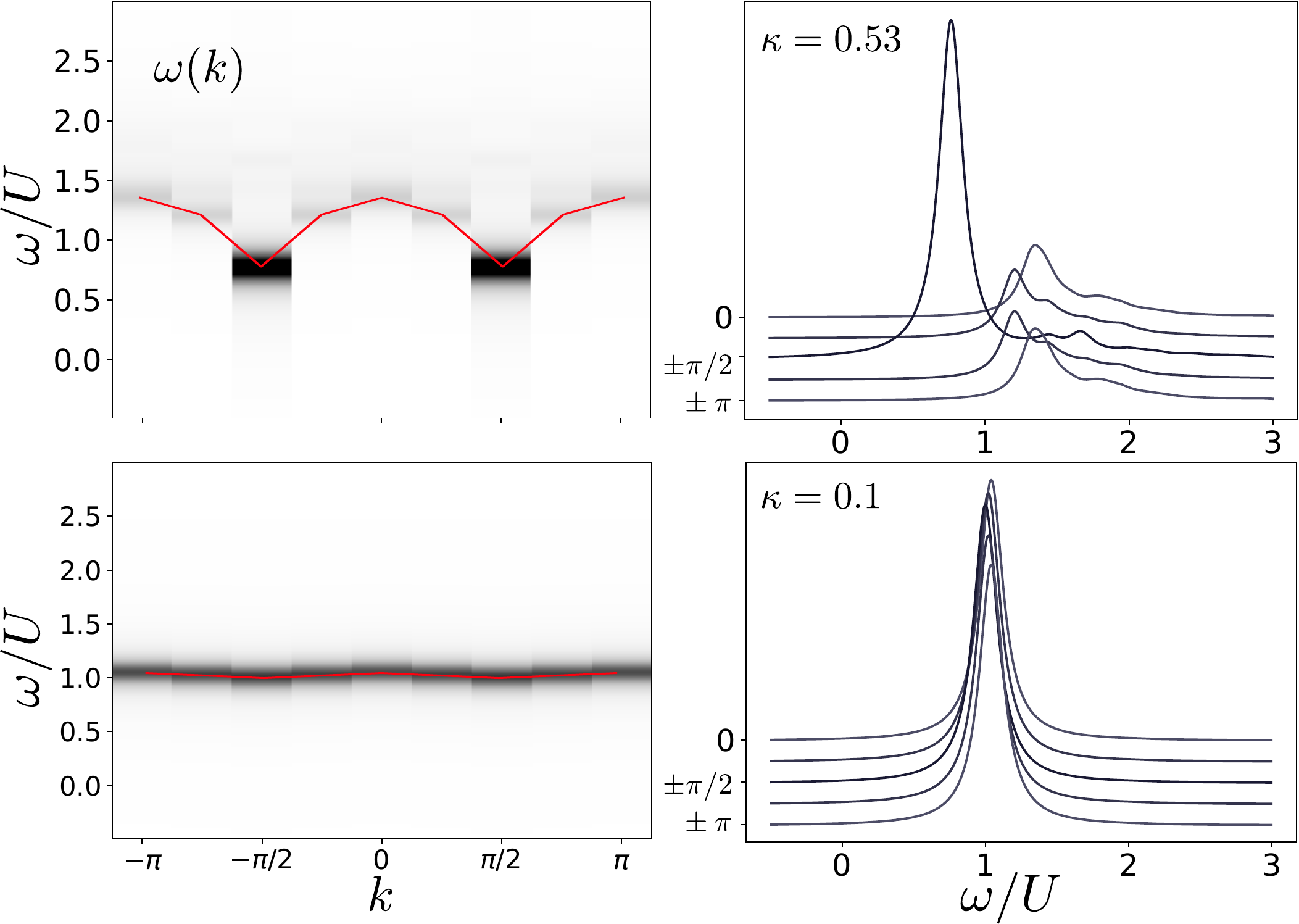}
\end{center}
\caption{Spectral density (left panels) and line shape (right panels) $A(\omega, k)$ of a single quasiparticle
excitation. The red lines in the left panels indicate the dispersion relation $\omega(q)$; the
velocity of excitations (the other Luttinger parameter) being given by
the slope of these curves at $k = \pm \pi/2$.
Upper plots are for the superfluid regime $\kappa = 0.53$, the lower panels for the Mott insulating regime $\kappa = 0.1$ for $N = 8$. }
\label{spec_dens}
\end{figure}

In the upper panels of Fig. \ref{spec_dens} we show the results obtained for a system well in the
superfluid regime ($\kappa = 0.53$). For each value of momentum the spectral function 
shows a single sharp peak, whose location shifts as the momentum changes.
The contour plot of $A(\omega, k)$ reveals that this momentum dependence
corresponds to two regions of linear dispersion, centered
on $k = \pm \pi/2$. We can thus
picture the ground state as consisting of two independent Luttinger liquids  
of bosons, condensed at these two different momenta but otherwise identical, with the
collective excitations being density fluctuations about these two condensates 
\footnote{By independent Luttinger liquids, we mean here two sets of bosons $A$ and $B$, occupying
different regions of the one-particle Hilbert space (here around momenta $\pm \pi/2$) such that
the total density fluctuation operator can be written as $\delta n_A (x) + \delta n_B (x)$
while ${\langle \Psi | \delta n_A (x) \delta n_B (x') | \Psi \rangle} = 0$.}.
In Appendix \ref{app_luttinger} we argue 
that $K_b=1/2$ remains the (commensurate case) critical value of this double Luttinger liquid system.

The lower panels of Fig. \ref{spec_dens} show the corresponding spectral functions when the system is
in the Mott insulator regime ($\kappa = 0.1$). In contrast to the previous result, the peaks
in $A(\omega, k)$ barely show any dependence on momentum, and the flat dispersion relation
obtained clearly indicates that the system has a gap of approximately $U$, exactly as would
be expected for a Mott insulator.

\section{\label{sec_scatter} Scattering processes}

As shown in  Figs. \ref{mom_den} and \ref{orbitals}, and discussed
in Section \ref{sec_results},
it is clear that the superfluid is
composed of a pair of macroscopically-occupied
degenerate orbitals, with momenta $\pm \pi/2$. The existence
of this fragmented condensate
has been explained on physical grounds
by noting that the dominant matrix elements linking the
Mott state and the dipole states correspond to single-particle hopping
processes with ``negative effective mass'' between next-nearest neighbor sites.
Here we examine the various
scattering amplitudes in more detail to try to understand the
dominant processes coexisting with the macroscopic occupation of $\pm \pi/2$ momentum eigenstates, and thus to obtain a microscopic picture
of the physics occurring.

An important piece of information that we can take from the simulations
of this system is that
the ground state only involves Fock states of zero total crystal momentum, 
i.e. the many-body ground state is an eigenstate of the total momentum with eigenvalue zero.
Quite generally, particles of initial momenta $k_l, k_m$ collide to final momenta $k_n,k_p$, with $k_n+k_p=k_l+k_m$, the connecting matrix elements
being given by Eq. \eqref{M-kappa} [see also Eq. \eqref{eq:Heff_mom_rep}]
The Bessel function appearing there can be written 
as ${\cal J}_0 \left( 2\kappa F \right)$, with
\begin{equation}
F= \cos k_l + \cos k_m - \cos k_n -\cos k_p   \, .
\label{eq:F-cos-k}
\end{equation}
Our goal here is to elucidate some basic features of the many-body ground state.

First we note that the sheer macroscopic occupation of a given momentum state does not favor a particular value of the privileged momentum, as $F$ is independent of $p$ when $k_l=k_m=k_n=k_p=p$. So the preference for a macroscopically occupied momentum will be determined by the ability of the condensate to connect with other configurations. More specifically, the success of a particular condensate choice will depend on how it relates to its depletion cloud. So the situation here is quite different from that of a standard condensate (such as the one yielded by the 
undriven BH model, for example). There the ground state is known to be dominated at zero temperature by the macroscopic occupation of the momentum $p=0$ combined with pair
collisions of the type $0,0 \leftrightarrow k,-k$, with $|k|$ small but nonzero. In the standard case, there are energetic reasons for favoring the macroscopic occupation of $p=0$, the depletion cloud merely contributing to the robustness of the resulting superfluid.

By contrast, here the connection between the condensate and its depletion cloud is essential to decide the macroscopically occupied momenta. A reasonable criterion is that the occupation of a given one-particle momentum state is favored in the ground state if it can intervene in processes yielding large matrix elements, as a suitable choice of the relative sign between the intervening many-body configurations permits a strong energy decrease. For moderate $\kappa$ [such that $4 \kappa F \alt x_1$, with $x_1=2.404$ the first zero of ${\cal J}_0(x)$] this translates into small values of $F$.

A generic pair collision of the type
\begin{equation}
p,q \longleftrightarrow p-k,q+k \, ,
\label{generic-pair-collision}
\end{equation}
has an associated matrix element which is proportional to \footnote{Umklapp processes with final momenta $(p-k,q+k+2\pi)$ yield the same $F$ value.}
\begin{equation}
F=(\cos p + \cos q)(1-\cos k)+(\sin p - \sin q)\sin k \, . 
\label{generic-F}
\end{equation}
For a given momentum pair $(p,q)$, the transferred momentum $k$ spans 
symmetrically around $k=0$. Thus, on average, when determining the typical value of $F$ for given $(p,q)$, 
the term in \eqref{generic-F} proportional to $\sin k$ will tend to cancel out,
while the $\cos k$ term will average to a non-zero value.
According to this criterion, the favored momenta will be those which minimize $|\cos p + \cos q|$, which requires $\cos q = - \cos p$. If now we note that the magnitudes of the matrix elements of boson collisions are enhanced by the macroscopic occupation of a single momentum state (bosonic enhancement),
we conclude that the case $p=q$ must also be advantageous. These two favorable requirements (zero average $F$ and $p=q$) can be satisfied simultaneously only for $p=q=\pm \pi/2$, in accordance with the numerical findings.

Thus, both from numerical evidence and semiquantitative arguments, 
the ground state appears to be dominated by configurations connected to each other by
processes involving pair collisions of the type
\begin{equation}
\frac{\pi}{2},\frac{\pi}{2} \longleftrightarrow \frac{\pi}{2}+k,\frac{\pi}{2}-k \, ,
\label{pair-pi2-pi2}
\end{equation}
against the background of a macroscopically occupied $\pi/2$-momentum state,
and similarly for the condensate at momentum $-\pi/2$. In both cases, the main processes are of the type \eqref{pair-pi2-pi2} with small but nonzero values of $| k |$, all yielding small
values of $F$ in \eqref{eq:F-cos-k} and thus large matrix elements.

Therefore the picture is reinforced of two condensates at momenta $\pm \pi/2$ 
with depletion clouds packed near those two momenta. 
To the extent that the two depletion clouds can be viewed as peaked 
around momenta $\pm \pi/2$ and thus non-overlapping, it seems fair to speak 
of a fragmented condensate, as suggested by our numerical results on the 
reduced one-particle density matrix (see Fig. \ref{orbitals}). This is also 
consistent with our results for the two-particle reduced density-matrix,
shown in  Fig. \ref{2particle_RDM}.

The thermodynamic limit of Hamiltonian \eqref{eq:Heff_mom_rep} remains to be studied in greater detail. A plausible outcome is that, as the one-dimensional system becomes larger, the relative atom density of both depletion clouds grows logarithmically. Thus the two quasi-condensates shrink in relative size while their respective depletion clouds grow and thus overlap more strongly through pairs of the type shown in the r.h.s. of \eqref{pair-pi2-pi2} with $|k|$ comparable to $\pi/2$, the result being a peculiar one-dimensional system.

\section{\label{sec_expt} Experimental implementation}

\begin{figure}
\begin{center}
\includegraphics[width=0.40\textwidth,clip=true]{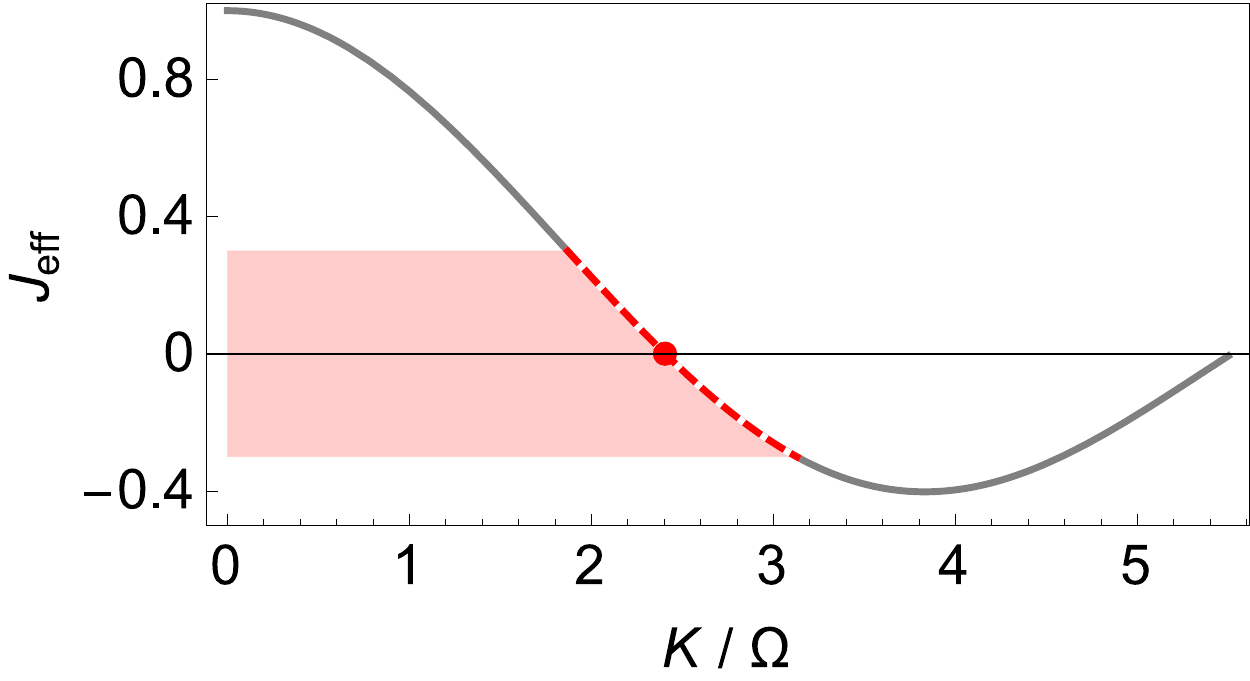}
\end{center}
\caption{The most frequently considered case of potential driving is
to ``shake'' the lattice potential sinusoidally with time. This produces an
effective tunneling $\Jeff = J {\cal J}_0(K / \Omega)$ where $\Omega$
is the frequency of the shaking and $K$ is its amplitude. If $K / \Omega$ is 
periodically varied about the first zero of the Bessel function,
$\Jeff$ will oscillate with same periodic time dependence in the shaded region,
yielding the Hamiltonian (\ref{eq:driven_hubb_ham}).}
\label{schematic}
\end{figure}

As we noted earlier, the form of driving that we choose is rather unusual.
In an optical lattice system it is straightforward to modulate
the depth of the system to alter the amplitude of the hopping
term, but we also require changing its {\em sign} so that its
time-average vanishes. One method of achieving this would be
to make use of the well-known lattice shaking technique,
which indeed provides a means of precisely controlling
the tunneling in this way. If a system of cold atoms,
described by a lattice tunneling model, is
periodically accelerated in space at a high shaking frequency $\Omega$, 
then the tunneling term is renormalized to an effective value. For sinusoidal
driving, for example, $\Jeff = J {\cal J}_0 \left( K / \Omega \right)$,
where $K$ is the amplitude of the shaking and ${\cal J}_0$ is the zeroth
Bessel function. We show this behaviour in Fig. \ref{schematic}, and
we can see that when the Bessel function is zero 
(at $K / \Omega \simeq 2.404$) the effective tunneling vanishes.

We can now consider altering the parameters of the shaking 
\cite{parametric_floquet} at a timescale much slower than
the period of the shaking $\tau = 2 \pi / \Omega$. The effective hopping 
will now vary according to $K(t) / \Omega$. In particular we can slowly 
oscillate $K$ about the root of the Bessel function
at a frequency $\omega$, where $\omega \ll \Omega$, which will 
give a time-dependent effective tunneling $\Jeff(t)$. For
a suitable choice of the variation $K(t)$, this will thus
produce the time-dependent
Hamiltonian (\ref{eq:driven_hubb_ham}). In order for 
Eq. \eqref{eq:driven_hubb_ham} to be in the high-frequency regime itself,
which we require for our analysis, we therefore need the driving
frequencies to follow the hierarchy: $J \ll \omega \ll \Omega$.

These conditions can be met in systems of periodically-driven
cold atoms such as rubidium-87 \cite{lignier2007dynamical,struck2012tunable}
and potassium-39 \cite{bloch_potassium}. In order to obtain a single-band
BH model as a starting point, it is necessary to use
optical lattice depths of $V_0 > 6 E_r$, where $E_r$ is the recoil
energy, in order to eliminate all except the nearest neighbor hopping terms.
This permits shaking frequencies of several tens of kilohertz to be
used without driving atoms into the next highest band, while the
bare tunneling, $J$, is of the order 100 Hz. This amply satisfies
the first condition, $J \ll \Omega$, and gives a wide window of available
frequencies for the frequency $\omega$ at which the tunneling is modulated.

A further consideration is that one might expect a periodically-driven
system to absorb energy from the driving, and thus eventually heat
up to an ``infinite temperature'', a phenomenon known as the eigenstate
thermalization hypothesis. If such heating occurs, it would be doubtful that the coherent
quantum effects we have studied would persist.
However, in the high-frequency limit, when the
driving frequency is much larger than the energy scales of the undriven
Hamiltonian, it has been shown \cite{abanin_2015, abanin_2017}
that a generic Floquet system first passes through a long-lived prethermal regime,
in which the heating rate is an exponentially small function of the driving frequency. 
Thermalization will eventually occur, but on a much longer timescale. 
In the prethermal regime,
the Floquet states of the system can be expected to have long lifetimes,
and the dynamics of the system can be well-approximated by
time evolution under the effective (Floquet) Hamiltonian.
As the Bose-Hubbard model is known to exhibit this behaviour \cite{bukov_2015},
and the driving we consider is well inside the high-frequency regime, 
we can be confident that the prethermal regime will be sufficiently large
to allow the unusual effects we predict to be experimentally observed.

Standard time-of-flight techniques can be used to image the momentum
density function, which should reveal the signatures of the fragmented
condensate we predict, shown in Fig. \ref{mom_den}.
While fragmented condensates can be expected to be unstable to perturbations
coupling the two fragments, we have verified that the state we observe
is robust to the presence of weak diagonal impurities.
Finally we have also verified that the effects we observe are not destroyed
by the presence of a residual static hopping, $J_\mathrm{dc}$, between
lattice sites, so long as $J_\mathrm{dc} / U$ is lower than the
critical value for the standard Mott transition. Consequently
we believe that its experimental realization is completely feasible.

\section{Conclusions}

We have shown that by driving the tunneling of the BH model with a zero average value, it is possible to produce an effective time-independent Hamiltonian with a number of surprising properties, a remarkable one being the macroscopic occupation, in the ground state, of momentum eigenstates $\pm \pi/2$.
While the creation of similar condensates has been
been theoretically predicted \cite{muramatsu2004,muramatsu2006}
and experimentally observed \cite{schneider} in the context of the
expansion of an initially confined Mott state, the effect
we report is rather different, being an {\em equilibrium} property of the many-body ground state. 
The sensitivity of the physics discussed here to the presence of impurities and the choice of boundary conditions
will be the subject of a future study.
The form of Floquet engineering investigated in this paper opens the prospect of
a new way of controlling the coherent dynamics of quantum systems,
and producing unusual states of matter.
Extending this work to higher dimensions and to fermionic systems
remain fascinating subjects for future research.

\acknowledgments
We would like to thank Wolfgang
Ketterle and Germ\'an Sierra for valuable discussions.
This work has been supported by Spain's MINECO through
Grant Nos. FIS2013-41716-P and
FIS2017-84368-P. One of us (FS) would like to acknowledge the 
support of the Real Colegio Complutense at Harvard and the Harvard-MIT 
Center for Ultracold Atoms, where part of this work was done.

\bibliographystyle{aipnum4-1}
\bibliography{j_paper_bib}

\begin{thebibliography}{61}%
\makeatletter
\providecommand \@ifxundefined [1]{%
 \@ifx{#1\undefined}
}%
\providecommand \@ifnum [1]{%
 \ifnum #1\expandafter \@firstoftwo
 \else \expandafter \@secondoftwo
 \fi
}%
\providecommand \@ifx [1]{%
 \ifx #1\expandafter \@firstoftwo
 \else \expandafter \@secondoftwo
 \fi
}%
\providecommand \natexlab [1]{#1}%
\providecommand \enquote  [1]{``#1''}%
\providecommand \bibnamefont  [1]{#1}%
\providecommand \bibfnamefont [1]{#1}%
\providecommand \citenamefont [1]{#1}%
\providecommand \href@noop [0]{\@secondoftwo}%
\providecommand \href [0]{\begingroup \@sanitize@url \@href}%
\providecommand \@href[1]{\@@startlink{#1}\@@href}%
\providecommand \@@href[1]{\endgroup#1\@@endlink}%
\providecommand \@sanitize@url [0]{\catcode `\\12\catcode `\$12\catcode
  `\&12\catcode `\#12\catcode `\^12\catcode `\_12\catcode `\%12\relax}%
\providecommand \@@startlink[1]{}%
\providecommand \@@endlink[0]{}%
\providecommand \url  [0]{\begingroup\@sanitize@url \@url }%
\providecommand \@url [1]{\endgroup\@href {#1}{\urlprefix }}%
\providecommand \urlprefix  [0]{URL }%
\providecommand \Eprint [0]{\href }%
\providecommand \doibase [0]{http://dx.doi.org/}%
\providecommand \selectlanguage [0]{\@gobble}%
\providecommand \bibinfo  [0]{\@secondoftwo}%
\providecommand \bibfield  [0]{\@secondoftwo}%
\providecommand \translation [1]{[#1]}%
\providecommand \BibitemOpen [0]{}%
\providecommand \bibitemStop [0]{}%
\providecommand \bibitemNoStop [0]{.\EOS\space}%
\providecommand \EOS [0]{\spacefactor3000\relax}%
\providecommand \BibitemShut  [1]{\csname bibitem#1\endcsname}%
\let\auto@bib@innerbib\@empty
%</preamble>
\bibitem [{\citenamefont {Eckardt}(2017)}]{eckardt_review}%
  \BibitemOpen
  \bibfield  {author} {\bibinfo {author} {\bibfnamefont {A.}~\bibnamefont
  {Eckardt}},\ }\href {\doibase 10.1103/RevModPhys.89.011004} {\bibfield
  {journal} {\bibinfo  {journal} {Rev. Mod. Phys.}\ }\textbf {\bibinfo {volume}
  {89}},\ \bibinfo {pages} {011004} (\bibinfo {year} {2017})}\BibitemShut
  {NoStop}%
\bibitem [{\citenamefont {Feynman}(1982)}]{feynman1982simulating}%
  \BibitemOpen
  \bibfield  {author} {\bibinfo {author} {\bibfnamefont {R.~P.}\ \bibnamefont
  {Feynman}},\ }\href {\doibase 10.1007/BF02650179} {\bibfield  {journal}
  {\bibinfo  {journal} {Int. J. Theor. Phys.}\ }\textbf {\bibinfo {volume}
  {21}},\ \bibinfo {pages} {467} (\bibinfo {year} {1982})}\BibitemShut
  {NoStop}%
\bibitem [{\citenamefont {Jaksch}\ and\ \citenamefont {Zoller}(2005)}]{jaksch}%
  \BibitemOpen
  \bibfield  {author} {\bibinfo {author} {\bibfnamefont {D.}~\bibnamefont
  {Jaksch}}\ and\ \bibinfo {author} {\bibfnamefont {P.}~\bibnamefont
  {Zoller}},\ }\href {\doibase 10.1016/j.aop.2004.09.010} {\bibfield  {journal}
  {\bibinfo  {journal} {Ann. Phys. (NY)}\ }\textbf {\bibinfo {volume} {315}},\
  \bibinfo {pages} {52} (\bibinfo {year} {2005})}\BibitemShut {NoStop}%
\bibitem [{\citenamefont {Lewenstein}\ \emph {et~al.}(2007)\citenamefont
  {Lewenstein}, \citenamefont {Sanpera}, \citenamefont {Ahufinger},
  \citenamefont {Damski}, \citenamefont {Sen},\ and\ \citenamefont
  {Sen}}]{lewenstein2007ultracold}%
  \BibitemOpen
  \bibfield  {author} {\bibinfo {author} {\bibfnamefont {M.}~\bibnamefont
  {Lewenstein}}, \bibinfo {author} {\bibfnamefont {A.}~\bibnamefont {Sanpera}},
  \bibinfo {author} {\bibfnamefont {V.}~\bibnamefont {Ahufinger}}, \bibinfo
  {author} {\bibfnamefont {B.}~\bibnamefont {Damski}}, \bibinfo {author}
  {\bibfnamefont {A.}~\bibnamefont {Sen}}, \ and\ \bibinfo {author}
  {\bibfnamefont {U.}~\bibnamefont {Sen}},\ }\href {\doibase
  10.1080/00018730701223200} {\bibfield  {journal} {\bibinfo  {journal} {Adv.
  Phys.}\ }\textbf {\bibinfo {volume} {56}},\ \bibinfo {pages} {243} (\bibinfo
  {year} {2007})}\BibitemShut {NoStop}%
\bibitem [{\citenamefont {Georgescu}, \citenamefont {Ashhab},\ and\
  \citenamefont {Nori}(2014)}]{georgescu2014quantum}%
  \BibitemOpen
  \bibfield  {author} {\bibinfo {author} {\bibfnamefont {I.}~\bibnamefont
  {Georgescu}}, \bibinfo {author} {\bibfnamefont {S.}~\bibnamefont {Ashhab}}, \
  and\ \bibinfo {author} {\bibfnamefont {F.}~\bibnamefont {Nori}},\ }\href
  {\doibase 10.1103/RevModPhys.86.153} {\bibfield  {journal} {\bibinfo
  {journal} {Rev. Mod. Phys.}\ }\textbf {\bibinfo {volume} {86}},\ \bibinfo
  {pages} {153} (\bibinfo {year} {2014})}\BibitemShut {NoStop}%
\bibitem [{\citenamefont {Creffield}\ and\ \citenamefont
  {Sierra}(2015)}]{creffield2015finding}%
  \BibitemOpen
  \bibfield  {author} {\bibinfo {author} {\bibfnamefont {C.~E.}\ \bibnamefont
  {Creffield}}\ and\ \bibinfo {author} {\bibfnamefont {G.}~\bibnamefont
  {Sierra}},\ }\href {\doibase 10.1103/PhysRevA.91.063608} {\bibfield
  {journal} {\bibinfo  {journal} {Phys. Rev. A}\ }\textbf {\bibinfo {volume}
  {91}},\ \bibinfo {pages} {063608} (\bibinfo {year} {2015})}\BibitemShut
  {NoStop}%
\bibitem [{\citenamefont {Iomin}\ and\ \citenamefont
  {Fishman}(2000)}]{Iomin2000}%
  \BibitemOpen
  \bibfield  {author} {\bibinfo {author} {\bibfnamefont {A.}~\bibnamefont
  {Iomin}}\ and\ \bibinfo {author} {\bibfnamefont {S.}~\bibnamefont
  {Fishman}},\ }\href {\doibase 10.1103/PhysRevB.61.2085} {\bibfield  {journal}
  {\bibinfo  {journal} {Phys. Rev. B}\ }\textbf {\bibinfo {volume} {61}},\
  \bibinfo {pages} {2085} (\bibinfo {year} {2000})}\BibitemShut {NoStop}%
\bibitem [{\citenamefont {Lindner}, \citenamefont {Refael},\ and\ \citenamefont
  {Galitski}(2011)}]{lindner2011floquet}%
  \BibitemOpen
  \bibfield  {author} {\bibinfo {author} {\bibfnamefont {N.~H.}\ \bibnamefont
  {Lindner}}, \bibinfo {author} {\bibfnamefont {G.}~\bibnamefont {Refael}}, \
  and\ \bibinfo {author} {\bibfnamefont {V.}~\bibnamefont {Galitski}},\ }\href
  {\doibase 10.1038/nphys1926} {\bibfield  {journal} {\bibinfo  {journal} {Nat.
  Phys.}\ }\textbf {\bibinfo {volume} {7}},\ \bibinfo {pages} {490} (\bibinfo
  {year} {2011})}\BibitemShut {NoStop}%
\bibitem [{\citenamefont {Zhang}\ \emph {et~al.}(2017)\citenamefont {Zhang},
  \citenamefont {Hess}, \citenamefont {Kyprianidis}, \citenamefont {Becker},
  \citenamefont {Lee}, \citenamefont {Smith}, \citenamefont {Pagano},
  \citenamefont {Potirniche}, \citenamefont {Potter}, \citenamefont
  {Vishwanath}, \citenamefont {Yao},\ and\ \citenamefont
  {Monroe}}]{time_crystals}%
  \BibitemOpen
  \bibfield  {author} {\bibinfo {author} {\bibfnamefont {J.}~\bibnamefont
  {Zhang}}, \bibinfo {author} {\bibfnamefont {P.~W.}\ \bibnamefont {Hess}},
  \bibinfo {author} {\bibfnamefont {A.}~\bibnamefont {Kyprianidis}}, \bibinfo
  {author} {\bibfnamefont {P.}~\bibnamefont {Becker}}, \bibinfo {author}
  {\bibfnamefont {A.}~\bibnamefont {Lee}}, \bibinfo {author} {\bibfnamefont
  {J.}~\bibnamefont {Smith}}, \bibinfo {author} {\bibfnamefont
  {G.}~\bibnamefont {Pagano}}, \bibinfo {author} {\bibfnamefont {I.~D.}\
  \bibnamefont {Potirniche}}, \bibinfo {author} {\bibfnamefont {A.~C.}\
  \bibnamefont {Potter}}, \bibinfo {author} {\bibfnamefont {A.}~\bibnamefont
  {Vishwanath}}, \bibinfo {author} {\bibfnamefont {N.~Y.}\ \bibnamefont {Yao}},
  \ and\ \bibinfo {author} {\bibfnamefont {C.}~\bibnamefont {Monroe}},\ }\href
  {\doibase 10.1038/nature21413} {\bibfield  {journal} {\bibinfo  {journal}
  {Nature}\ }\textbf {\bibinfo {volume} {543}},\ \bibinfo {pages} {217}
  (\bibinfo {year} {2017})}\BibitemShut {NoStop}%
\bibitem [{\citenamefont {Choi}\ \emph {et~al.}(2017)\citenamefont {Choi},
  \citenamefont {Choi}, \citenamefont {Landig}, \citenamefont {Kucsko},
  \citenamefont {Zhou}, \citenamefont {Isoya}, \citenamefont {Jelezko},
  \citenamefont {Onoda}, \citenamefont {Sumiya}, \citenamefont {Khemani},
  \citenamefont {von Keyserlingk}, \citenamefont {Yao}, \citenamefont
  {Demler},\ and\ \citenamefont {Lukin}}]{time_crystals2}%
  \BibitemOpen
  \bibfield  {author} {\bibinfo {author} {\bibfnamefont {S.}~\bibnamefont
  {Choi}}, \bibinfo {author} {\bibfnamefont {J.}~\bibnamefont {Choi}}, \bibinfo
  {author} {\bibfnamefont {R.}~\bibnamefont {Landig}}, \bibinfo {author}
  {\bibfnamefont {G.}~\bibnamefont {Kucsko}}, \bibinfo {author} {\bibfnamefont
  {H.}~\bibnamefont {Zhou}}, \bibinfo {author} {\bibfnamefont {J.}~\bibnamefont
  {Isoya}}, \bibinfo {author} {\bibfnamefont {F.}~\bibnamefont {Jelezko}},
  \bibinfo {author} {\bibfnamefont {S.}~\bibnamefont {Onoda}}, \bibinfo
  {author} {\bibfnamefont {H.}~\bibnamefont {Sumiya}}, \bibinfo {author}
  {\bibfnamefont {V.}~\bibnamefont {Khemani}}, \bibinfo {author} {\bibfnamefont
  {C.}~\bibnamefont {von Keyserlingk}}, \bibinfo {author} {\bibfnamefont
  {N.~Y.}\ \bibnamefont {Yao}}, \bibinfo {author} {\bibfnamefont
  {E.}~\bibnamefont {Demler}}, \ and\ \bibinfo {author} {\bibfnamefont {M.~D.}\
  \bibnamefont {Lukin}},\ }\href {\doibase 10.1038/nature21426} {\bibfield
  {journal} {\bibinfo  {journal} {Nature}\ }\textbf {\bibinfo {volume} {543}},\
  \bibinfo {pages} {221–225} (\bibinfo {year} {2017})}\BibitemShut {NoStop}%
\bibitem [{\citenamefont {Dunlap}\ and\ \citenamefont {Kenkre}(1986)}]{dunlap}%
  \BibitemOpen
  \bibfield  {author} {\bibinfo {author} {\bibfnamefont {D.~H.}\ \bibnamefont
  {Dunlap}}\ and\ \bibinfo {author} {\bibfnamefont {V.~M.}\ \bibnamefont
  {Kenkre}},\ }\href {\doibase 10.1103/PhysRevB.34.3625} {\bibfield  {journal}
  {\bibinfo  {journal} {Phys. Rev. B}\ }\textbf {\bibinfo {volume} {34}},\
  \bibinfo {pages} {3625} (\bibinfo {year} {1986})}\BibitemShut {NoStop}%
\bibitem [{\citenamefont {Grifoni}\ and\ \citenamefont
  {H{\"a}nggi}(1998)}]{cdt}%
  \BibitemOpen
  \bibfield  {author} {\bibinfo {author} {\bibfnamefont {M.}~\bibnamefont
  {Grifoni}}\ and\ \bibinfo {author} {\bibfnamefont {P.}~\bibnamefont
  {H{\"a}nggi}},\ }\href {\doibase 10.1016/S0370-1573(98)00022-2} {\bibfield
  {journal} {\bibinfo  {journal} {Phys. Rep.}\ }\textbf {\bibinfo {volume}
  {304}},\ \bibinfo {pages} {229} (\bibinfo {year} {1998})}\BibitemShut
  {NoStop}%
\bibitem [{\citenamefont {Lignier}\ \emph {et~al.}(2007)\citenamefont
  {Lignier}, \citenamefont {Sias}, \citenamefont {Ciampini}, \citenamefont
  {Singh}, \citenamefont {Zenesini}, \citenamefont {Morsch},\ and\
  \citenamefont {Arimondo}}]{lignier2007dynamical}%
  \BibitemOpen
  \bibfield  {author} {\bibinfo {author} {\bibfnamefont {H.}~\bibnamefont
  {Lignier}}, \bibinfo {author} {\bibfnamefont {C.}~\bibnamefont {Sias}},
  \bibinfo {author} {\bibfnamefont {D.}~\bibnamefont {Ciampini}}, \bibinfo
  {author} {\bibfnamefont {Y.}~\bibnamefont {Singh}}, \bibinfo {author}
  {\bibfnamefont {A.}~\bibnamefont {Zenesini}}, \bibinfo {author}
  {\bibfnamefont {O.}~\bibnamefont {Morsch}}, \ and\ \bibinfo {author}
  {\bibfnamefont {E.}~\bibnamefont {Arimondo}},\ }\href {\doibase
  10.1103/PhysRevLett.99.220403} {\bibfield  {journal} {\bibinfo  {journal}
  {Phys. Rev. Lett.}\ }\textbf {\bibinfo {volume} {99}},\ \bibinfo {pages}
  {220403} (\bibinfo {year} {2007})}\BibitemShut {NoStop}%
\bibitem [{\citenamefont {Kierig}\ \emph {et~al.}(2008)\citenamefont {Kierig},
  \citenamefont {Schnorrberger}, \citenamefont {Schietinger}, \citenamefont
  {Tomkovic},\ and\ \citenamefont {Oberthaler}}]{kierig2008single}%
  \BibitemOpen
  \bibfield  {author} {\bibinfo {author} {\bibfnamefont {E.}~\bibnamefont
  {Kierig}}, \bibinfo {author} {\bibfnamefont {U.}~\bibnamefont
  {Schnorrberger}}, \bibinfo {author} {\bibfnamefont {A.}~\bibnamefont
  {Schietinger}}, \bibinfo {author} {\bibfnamefont {J.}~\bibnamefont
  {Tomkovic}}, \ and\ \bibinfo {author} {\bibfnamefont {M.}~\bibnamefont
  {Oberthaler}},\ }\href {\doibase 10.1103/PhysRevLett.100.190405} {\bibfield
  {journal} {\bibinfo  {journal} {Phys. Rev. Lett.}\ }\textbf {\bibinfo
  {volume} {100}},\ \bibinfo {pages} {190405} (\bibinfo {year}
  {2008})}\BibitemShut {NoStop}%
\bibitem [{\citenamefont {Eckardt}, \citenamefont {Weiss},\ and\ \citenamefont
  {Holthaus}(2005)}]{eckardt}%
  \BibitemOpen
  \bibfield  {author} {\bibinfo {author} {\bibfnamefont {A.}~\bibnamefont
  {Eckardt}}, \bibinfo {author} {\bibfnamefont {C.}~\bibnamefont {Weiss}}, \
  and\ \bibinfo {author} {\bibfnamefont {M.}~\bibnamefont {Holthaus}},\ }\href
  {\doibase 10.1103/PhysRevLett.95.260404} {\bibfield  {journal} {\bibinfo
  {journal} {Phys. Rev. Lett.}\ }\textbf {\bibinfo {volume} {95}},\ \bibinfo
  {pages} {260404} (\bibinfo {year} {2005})}\BibitemShut {NoStop}%
\bibitem [{\citenamefont {Creffield}\ and\ \citenamefont
  {Monteiro}(2006)}]{cec_tsm}%
  \BibitemOpen
  \bibfield  {author} {\bibinfo {author} {\bibfnamefont {C.~E.}\ \bibnamefont
  {Creffield}}\ and\ \bibinfo {author} {\bibfnamefont {T.~S.}\ \bibnamefont
  {Monteiro}},\ }\href {\doibase 10.1103/PhysRevLett.96.210403} {\bibfield
  {journal} {\bibinfo  {journal} {Phys. Rev. Lett.}\ }\textbf {\bibinfo
  {volume} {96}},\ \bibinfo {pages} {210403} (\bibinfo {year}
  {2006})}\BibitemShut {NoStop}%
\bibitem [{\citenamefont {Zenesini}\ \emph {et~al.}(2009)\citenamefont
  {Zenesini}, \citenamefont {Lignier}, \citenamefont {Ciampini}, \citenamefont
  {Morsch},\ and\ \citenamefont {Arimondo}}]{zenesini2009parametric}%
  \BibitemOpen
  \bibfield  {author} {\bibinfo {author} {\bibfnamefont {A.}~\bibnamefont
  {Zenesini}}, \bibinfo {author} {\bibfnamefont {H.}~\bibnamefont {Lignier}},
  \bibinfo {author} {\bibfnamefont {D.}~\bibnamefont {Ciampini}}, \bibinfo
  {author} {\bibfnamefont {O.}~\bibnamefont {Morsch}}, \ and\ \bibinfo {author}
  {\bibfnamefont {E.}~\bibnamefont {Arimondo}},\ }\href {\doibase
  10.1103/PhysRevLett.102.100403} {\bibfield  {journal} {\bibinfo  {journal}
  {Phys. Rev. Lett.}\ }\textbf {\bibinfo {volume} {102}},\ \bibinfo {pages}
  {100403} (\bibinfo {year} {2009})}\BibitemShut {NoStop}%
\bibitem [{\citenamefont {Kolovsky}(2011)}]{kolovsky2011creating}%
  \BibitemOpen
  \bibfield  {author} {\bibinfo {author} {\bibfnamefont {A.~R.}\ \bibnamefont
  {Kolovsky}},\ }\href {\doibase 10.1209/0295-5075/93/20003} {\bibfield
  {journal} {\bibinfo  {journal} {EPL}\ }\textbf {\bibinfo {volume} {93}},\
  \bibinfo {pages} {20003} (\bibinfo {year} {2011})}\BibitemShut {NoStop}%
\bibitem [{\citenamefont {Creffield}\ and\ \citenamefont
  {Sols}(2013)}]{creffield2013comment}%
  \BibitemOpen
  \bibfield  {author} {\bibinfo {author} {\bibfnamefont {C.~E.}\ \bibnamefont
  {Creffield}}\ and\ \bibinfo {author} {\bibfnamefont {F.}~\bibnamefont
  {Sols}},\ }\href {\doibase 10.1209/0295-5075/101/40001} {\bibfield  {journal}
  {\bibinfo  {journal} {EPL}\ }\textbf {\bibinfo {volume} {101}},\ \bibinfo
  {pages} {40001} (\bibinfo {year} {2013})}\BibitemShut {NoStop}%
\bibitem [{\citenamefont {Struck}\ \emph {et~al.}(2012)\citenamefont {Struck},
  \citenamefont {{\"O}lschl{\"a}ger}, \citenamefont {Weinberg}, \citenamefont
  {Hauke}, \citenamefont {Simonet}, \citenamefont {Eckardt}, \citenamefont
  {Lewenstein}, \citenamefont {Sengstock},\ and\ \citenamefont
  {Windpassinger}}]{struck2012tunable}%
  \BibitemOpen
  \bibfield  {author} {\bibinfo {author} {\bibfnamefont {J.}~\bibnamefont
  {Struck}}, \bibinfo {author} {\bibfnamefont {C.}~\bibnamefont
  {{\"O}lschl{\"a}ger}}, \bibinfo {author} {\bibfnamefont {M.}~\bibnamefont
  {Weinberg}}, \bibinfo {author} {\bibfnamefont {P.}~\bibnamefont {Hauke}},
  \bibinfo {author} {\bibfnamefont {J.}~\bibnamefont {Simonet}}, \bibinfo
  {author} {\bibfnamefont {A.}~\bibnamefont {Eckardt}}, \bibinfo {author}
  {\bibfnamefont {M.}~\bibnamefont {Lewenstein}}, \bibinfo {author}
  {\bibfnamefont {K.}~\bibnamefont {Sengstock}}, \ and\ \bibinfo {author}
  {\bibfnamefont {P.}~\bibnamefont {Windpassinger}},\ }\href {\doibase
  10.1103/PhysRevLett.108.225304} {\bibfield  {journal} {\bibinfo  {journal}
  {Phys. Rev. Lett.}\ }\textbf {\bibinfo {volume} {108}},\ \bibinfo {pages}
  {225304} (\bibinfo {year} {2012})}\BibitemShut {NoStop}%
\bibitem [{\citenamefont {Hauke}\ \emph {et~al.}(2012)\citenamefont {Hauke},
  \citenamefont {Tieleman}, \citenamefont {Celi}, \citenamefont
  {\"Olschl\"ager}, \citenamefont {Simonet}, \citenamefont {Struck},
  \citenamefont {Weinberg}, \citenamefont {Windpassinger}, \citenamefont
  {Sengstock}, \citenamefont {Lewenstein},\ and\ \citenamefont
  {Eckardt}}]{struck2}%
  \BibitemOpen
  \bibfield  {author} {\bibinfo {author} {\bibfnamefont {P.}~\bibnamefont
  {Hauke}}, \bibinfo {author} {\bibfnamefont {O.}~\bibnamefont {Tieleman}},
  \bibinfo {author} {\bibfnamefont {A.}~\bibnamefont {Celi}}, \bibinfo {author}
  {\bibfnamefont {C.}~\bibnamefont {\"Olschl\"ager}}, \bibinfo {author}
  {\bibfnamefont {J.}~\bibnamefont {Simonet}}, \bibinfo {author} {\bibfnamefont
  {J.}~\bibnamefont {Struck}}, \bibinfo {author} {\bibfnamefont
  {M.}~\bibnamefont {Weinberg}}, \bibinfo {author} {\bibfnamefont
  {P.}~\bibnamefont {Windpassinger}}, \bibinfo {author} {\bibfnamefont
  {K.}~\bibnamefont {Sengstock}}, \bibinfo {author} {\bibfnamefont
  {M.}~\bibnamefont {Lewenstein}}, \ and\ \bibinfo {author} {\bibfnamefont
  {A.}~\bibnamefont {Eckardt}},\ }\href {\doibase
  10.1103/PhysRevLett.109.145301} {\bibfield  {journal} {\bibinfo  {journal}
  {Phys. Rev. Lett.}\ }\textbf {\bibinfo {volume} {109}},\ \bibinfo {pages}
  {145301} (\bibinfo {year} {2012})}\BibitemShut {NoStop}%
\bibitem [{\citenamefont {Rapp}, \citenamefont {Deng},\ and\ \citenamefont
  {Santos}(2012)}]{santos}%
  \BibitemOpen
  \bibfield  {author} {\bibinfo {author} {\bibfnamefont {{\'A}.}~\bibnamefont
  {Rapp}}, \bibinfo {author} {\bibfnamefont {X.}~\bibnamefont {Deng}}, \ and\
  \bibinfo {author} {\bibfnamefont {L.}~\bibnamefont {Santos}},\ }\href
  {\doibase 10.1103/PhysRevLett.109.203005} {\bibfield  {journal} {\bibinfo
  {journal} {Phys. Rev. Lett.}\ }\textbf {\bibinfo {volume} {109}},\ \bibinfo
  {pages} {203005} (\bibinfo {year} {2012})}\BibitemShut {NoStop}%
\bibitem [{\citenamefont {Di~Liberto}\ \emph {et~al.}(2014)\citenamefont
  {Di~Liberto}, \citenamefont {Creffield}, \citenamefont {Japaridze},\ and\
  \citenamefont {Smith}}]{liberto}%
  \BibitemOpen
  \bibfield  {author} {\bibinfo {author} {\bibfnamefont {M.}~\bibnamefont
  {Di~Liberto}}, \bibinfo {author} {\bibfnamefont {C.~E.}\ \bibnamefont
  {Creffield}}, \bibinfo {author} {\bibfnamefont {G.}~\bibnamefont
  {Japaridze}}, \ and\ \bibinfo {author} {\bibfnamefont {C.~M.}\ \bibnamefont
  {Smith}},\ }\href {\doibase 10.1103/PhysRevA.89.013624} {\bibfield  {journal}
  {\bibinfo  {journal} {Phys. Rev. A}\ }\textbf {\bibinfo {volume} {89}},\
  \bibinfo {pages} {013624} (\bibinfo {year} {2014})}\BibitemShut {NoStop}%
\bibitem [{\citenamefont {Gaul}\ \emph {et~al.}(2011)\citenamefont {Gaul},
  \citenamefont {D{\'\i}az}, \citenamefont {Lima}, \citenamefont
  {Dom{\'\i}nguez-Adame},\ and\ \citenamefont
  {M{\"u}ller}}]{gaul2011stability}%
  \BibitemOpen
  \bibfield  {author} {\bibinfo {author} {\bibfnamefont {C.}~\bibnamefont
  {Gaul}}, \bibinfo {author} {\bibfnamefont {E.}~\bibnamefont {D{\'\i}az}},
  \bibinfo {author} {\bibfnamefont {R.~P.}\ \bibnamefont {Lima}}, \bibinfo
  {author} {\bibfnamefont {F.}~\bibnamefont {Dom{\'\i}nguez-Adame}}, \ and\
  \bibinfo {author} {\bibfnamefont {C.~A.}\ \bibnamefont {M{\"u}ller}},\ }\href
  {\doibase 10.1103/PhysRevA.84.053627} {\bibfield  {journal} {\bibinfo
  {journal} {Phys. Rev. A}\ }\textbf {\bibinfo {volume} {84}},\ \bibinfo
  {pages} {053627} (\bibinfo {year} {2011})}\BibitemShut {NoStop}%
\bibitem [{\citenamefont {Meinert}\ \emph {et~al.}(2016)\citenamefont
  {Meinert}, \citenamefont {Mark}, \citenamefont {Lauber}, \citenamefont
  {Daley},\ and\ \citenamefont {N{\"a}gerl}}]{meinert2016floquet}%
  \BibitemOpen
  \bibfield  {author} {\bibinfo {author} {\bibfnamefont {F.}~\bibnamefont
  {Meinert}}, \bibinfo {author} {\bibfnamefont {M.~J.}\ \bibnamefont {Mark}},
  \bibinfo {author} {\bibfnamefont {K.}~\bibnamefont {Lauber}}, \bibinfo
  {author} {\bibfnamefont {A.~J.}\ \bibnamefont {Daley}}, \ and\ \bibinfo
  {author} {\bibfnamefont {H.-C.}\ \bibnamefont {N{\"a}gerl}},\ }\href
  {\doibase 10.1103/PhysRevLett.116.205301} {\bibfield  {journal} {\bibinfo
  {journal} {Phys. Rev. Lett.}\ }\textbf {\bibinfo {volume} {116}},\ \bibinfo
  {pages} {205301} (\bibinfo {year} {2016})}\BibitemShut {NoStop}%
\bibitem [{\citenamefont {Dutta}\ \emph {et~al.}(2015)\citenamefont {Dutta},
  \citenamefont {Gajda}, \citenamefont {Hauke}, \citenamefont {Lewenstein},
  \citenamefont {L{\"u}hmann}, \citenamefont {Malomed}, \citenamefont
  {Sowi{\'n}ski},\ and\ \citenamefont {Zakrzewski}}]{nonstandard}%
  \BibitemOpen
  \bibfield  {author} {\bibinfo {author} {\bibfnamefont {O.}~\bibnamefont
  {Dutta}}, \bibinfo {author} {\bibfnamefont {M.}~\bibnamefont {Gajda}},
  \bibinfo {author} {\bibfnamefont {P.}~\bibnamefont {Hauke}}, \bibinfo
  {author} {\bibfnamefont {M.}~\bibnamefont {Lewenstein}}, \bibinfo {author}
  {\bibfnamefont {D.-S.}\ \bibnamefont {L{\"u}hmann}}, \bibinfo {author}
  {\bibfnamefont {B.~A.}\ \bibnamefont {Malomed}}, \bibinfo {author}
  {\bibfnamefont {T.}~\bibnamefont {Sowi{\'n}ski}}, \ and\ \bibinfo {author}
  {\bibfnamefont {J.}~\bibnamefont {Zakrzewski}},\ }\href
  {http://stacks.iop.org/0034-4885/78/i=6/a=066001} {\bibfield  {journal}
  {\bibinfo  {journal} {Reports on Progress in Physics}\ }\textbf {\bibinfo
  {volume} {78}},\ \bibinfo {pages} {066001} (\bibinfo {year}
  {2015})}\BibitemShut {NoStop}%
\bibitem [{\citenamefont {Str\"ater}, \citenamefont {Srivastava},\ and\
  \citenamefont {Eckardt}(2016)}]{anyon}%
  \BibitemOpen
  \bibfield  {author} {\bibinfo {author} {\bibfnamefont {C.}~\bibnamefont
  {Str\"ater}}, \bibinfo {author} {\bibfnamefont {S.~C.~L.}\ \bibnamefont
  {Srivastava}}, \ and\ \bibinfo {author} {\bibfnamefont {A.}~\bibnamefont
  {Eckardt}},\ }\href {\doibase 10.1103/PhysRevLett.117.205303} {\bibfield
  {journal} {\bibinfo  {journal} {Phys. Rev. Lett.}\ }\textbf {\bibinfo
  {volume} {117}},\ \bibinfo {pages} {205303} (\bibinfo {year}
  {2016})}\BibitemShut {NoStop}%
\bibitem [{\citenamefont {St\"oferle}\ \emph {et~al.}(2004)\citenamefont
  {St\"oferle}, \citenamefont {Moritz}, \citenamefont {Schori}, \citenamefont
  {K\"ohl},\ and\ \citenamefont {Esslinger}}]{stoeferl}%
  \BibitemOpen
  \bibfield  {author} {\bibinfo {author} {\bibfnamefont {T.}~\bibnamefont
  {St\"oferle}}, \bibinfo {author} {\bibfnamefont {H.}~\bibnamefont {Moritz}},
  \bibinfo {author} {\bibfnamefont {C.}~\bibnamefont {Schori}}, \bibinfo
  {author} {\bibfnamefont {M.}~\bibnamefont {K\"ohl}}, \ and\ \bibinfo {author}
  {\bibfnamefont {T.}~\bibnamefont {Esslinger}},\ }\href {\doibase
  10.1103/PhysRevLett.92.130403} {\bibfield  {journal} {\bibinfo  {journal}
  {Phys. Rev. Lett.}\ }\textbf {\bibinfo {volume} {92}},\ \bibinfo {pages}
  {130403} (\bibinfo {year} {2004})}\BibitemShut {NoStop}%
\bibitem [{\citenamefont {Kollath}\ \emph {et~al.}(2006)\citenamefont
  {Kollath}, \citenamefont {Iucci}, \citenamefont {Giamarchi}, \citenamefont
  {Hofstetter},\ and\ \citenamefont
  {Schollw{\"o}ck}}]{kollath2006spectroscopy}%
  \BibitemOpen
  \bibfield  {author} {\bibinfo {author} {\bibfnamefont {C.}~\bibnamefont
  {Kollath}}, \bibinfo {author} {\bibfnamefont {A.}~\bibnamefont {Iucci}},
  \bibinfo {author} {\bibfnamefont {T.}~\bibnamefont {Giamarchi}}, \bibinfo
  {author} {\bibfnamefont {W.}~\bibnamefont {Hofstetter}}, \ and\ \bibinfo
  {author} {\bibfnamefont {U.}~\bibnamefont {Schollw{\"o}ck}},\ }\href
  {\doibase 10.1103/PhysRevLett.97.050402} {\bibfield  {journal} {\bibinfo
  {journal} {Phys. Rev. Lett.}\ }\textbf {\bibinfo {volume} {97}},\ \bibinfo
  {pages} {050402} (\bibinfo {year} {2006})}\BibitemShut {NoStop}%
\bibitem [{\citenamefont {Dirks}\ \emph {et~al.}(2014)\citenamefont {Dirks},
  \citenamefont {Mikelsons}, \citenamefont {Krishnamurthy},\ and\ \citenamefont
  {Freericks}}]{dirks2014theoretical}%
  \BibitemOpen
  \bibfield  {author} {\bibinfo {author} {\bibfnamefont {A.}~\bibnamefont
  {Dirks}}, \bibinfo {author} {\bibfnamefont {K.}~\bibnamefont {Mikelsons}},
  \bibinfo {author} {\bibfnamefont {H.}~\bibnamefont {Krishnamurthy}}, \ and\
  \bibinfo {author} {\bibfnamefont {J.}~\bibnamefont {Freericks}},\ }\href
  {\doibase 10.1103/PhysRevA.89.021602} {\bibfield  {journal} {\bibinfo
  {journal} {Phys. Rev. A}\ }\textbf {\bibinfo {volume} {89}},\ \bibinfo
  {pages} {021602} (\bibinfo {year} {2014})}\BibitemShut {NoStop}%
\bibitem [{\citenamefont {Cardarelli}, \citenamefont {Greschner},\ and\
  \citenamefont {Santos}(2016)}]{cardarelli2016}%
  \BibitemOpen
  \bibfield  {author} {\bibinfo {author} {\bibfnamefont {L.}~\bibnamefont
  {Cardarelli}}, \bibinfo {author} {\bibfnamefont {S.}~\bibnamefont
  {Greschner}}, \ and\ \bibinfo {author} {\bibfnamefont {L.}~\bibnamefont
  {Santos}},\ }\href {\doibase 10.1103/PhysRevA.94.023615} {\bibfield
  {journal} {\bibinfo  {journal} {Phys. Rev. A}\ }\textbf {\bibinfo {volume}
  {94}},\ \bibinfo {pages} {023615} (\bibinfo {year} {2016})}\BibitemShut
  {NoStop}%
\bibitem [{\citenamefont {Sachdev}(2001)}]{sachdev}%
  \BibitemOpen
  \bibfield  {author} {\bibinfo {author} {\bibfnamefont {S.}~\bibnamefont
  {Sachdev}},\ }\href {https://books.google.co.uk/books?id=Ih\_E05N5TZQC}
  {\emph {\bibinfo {title} {Quantum Phase Transitions}}}\ (\bibinfo
  {publisher} {Cambridge University Press},\ \bibinfo {year}
  {2001})\BibitemShut {NoStop}%
\bibitem [{Note1()}]{Note1}%
  \BibitemOpen
  \bibinfo {note} {We have also considered squarewave and triangular driving
  functions, and find that they give very similar behaviour.}\BibitemShut
  {Stop}%
\bibitem [{\citenamefont {Holthaus}(2016)}]{holthaus_review}%
  \BibitemOpen
  \bibfield  {author} {\bibinfo {author} {\bibfnamefont {M.}~\bibnamefont
  {Holthaus}},\ }\href {http://stacks.iop.org/0953-4075/49/i=1/a=013001}
  {\bibfield  {journal} {\bibinfo  {journal} {Journal of Physics B: Atomic,
  Molecular and Optical Physics}\ }\textbf {\bibinfo {volume} {49}},\ \bibinfo
  {pages} {013001} (\bibinfo {year} {2016})}\BibitemShut {NoStop}%
\bibitem [{\citenamefont {Goldman}\ and\ \citenamefont
  {Dalibard}(2014)}]{goldman_prx}%
  \BibitemOpen
  \bibfield  {author} {\bibinfo {author} {\bibfnamefont {N.}~\bibnamefont
  {Goldman}}\ and\ \bibinfo {author} {\bibfnamefont {J.}~\bibnamefont
  {Dalibard}},\ }\href {\doibase 10.1103/PhysRevX.4.031027} {\bibfield
  {journal} {\bibinfo  {journal} {Phys. Rev. X}\ }\textbf {\bibinfo {volume}
  {4}},\ \bibinfo {pages} {031027} (\bibinfo {year} {2014})}\BibitemShut
  {NoStop}%
\bibitem [{\citenamefont {Creffield}\ \emph {et~al.}(2016)\citenamefont
  {Creffield}, \citenamefont {Pieplow}, \citenamefont {Sols},\ and\
  \citenamefont {Goldman}}]{creffield2016realization}%
  \BibitemOpen
  \bibfield  {author} {\bibinfo {author} {\bibfnamefont {C.~E.}\ \bibnamefont
  {Creffield}}, \bibinfo {author} {\bibfnamefont {G.}~\bibnamefont {Pieplow}},
  \bibinfo {author} {\bibfnamefont {F.}~\bibnamefont {Sols}}, \ and\ \bibinfo
  {author} {\bibfnamefont {N.}~\bibnamefont {Goldman}},\ }\href {\doibase
  10.1088/1367-2630/18/9/093013} {\bibfield  {journal} {\bibinfo  {journal}
  {New J. Phys.}\ }\textbf {\bibinfo {volume} {18}},\ \bibinfo {pages} {093013}
  (\bibinfo {year} {2016})}\BibitemShut {NoStop}%
\bibitem [{\citenamefont {Bukov}, \citenamefont {D'Alessio},\ and\
  \citenamefont {Polkovnikov}(2015)}]{bukov}%
  \BibitemOpen
  \bibfield  {author} {\bibinfo {author} {\bibfnamefont {M.}~\bibnamefont
  {Bukov}}, \bibinfo {author} {\bibfnamefont {L.}~\bibnamefont {D'Alessio}}, \
  and\ \bibinfo {author} {\bibfnamefont {A.}~\bibnamefont {Polkovnikov}},\
  }\href {\doibase 10.1080/00018732.2015.1055918} {\bibfield  {journal}
  {\bibinfo  {journal} {Advances in Physics}\ }\textbf {\bibinfo {volume}
  {64}},\ \bibinfo {pages} {139} (\bibinfo {year} {2015})},\ \Eprint
  {http://arxiv.org/abs/http://dx.doi.org/10.1080/00018732.2015.1055918}
  {http://dx.doi.org/10.1080/00018732.2015.1055918} \BibitemShut {NoStop}%
\bibitem [{\citenamefont {Eckardt}\ and\ \citenamefont
  {Anisimovas}(2015)}]{anisimovas}%
  \BibitemOpen
  \bibfield  {author} {\bibinfo {author} {\bibfnamefont {A.}~\bibnamefont
  {Eckardt}}\ and\ \bibinfo {author} {\bibfnamefont {E.}~\bibnamefont
  {Anisimovas}},\ }\href {http://stacks.iop.org/1367-2630/17/i=9/a=093039}
  {\bibfield  {journal} {\bibinfo  {journal} {New Journal of Physics}\ }\textbf
  {\bibinfo {volume} {17}},\ \bibinfo {pages} {093039} (\bibinfo {year}
  {2015})}\BibitemShut {NoStop}%
\bibitem [{Note2()}]{Note2}%
  \BibitemOpen
  \bibinfo {note} {Comparing the quasienergies $\epsilon _j$ and the
  eigenenergies of the effective Hamiltonian $E^\protect \mathrm {eff}_j$ by
  evaluating $\chi ^2 = \DOTSB \sum@ \slimits@ |\epsilon _j - E^\protect
  \mathrm {eff}_j |^2$, we find that for a fixed value of $\kappa $ it is bound
  by $\chi ^2 < 5\times 10^{-7}$. The ordering and the values of the
  quasienergies are thus essentially indistinguishable from the energy levels
  of the effective Hamiltonian.}\BibitemShut {Stop}%
\bibitem [{\citenamefont {Benito}\ \emph {et~al.}(2014)\citenamefont {Benito},
  \citenamefont {G{\'o}mez-Le{\'o}n}, \citenamefont {Bastidas}, \citenamefont
  {Brandes},\ and\ \citenamefont {Platero}}]{monica}%
  \BibitemOpen
  \bibfield  {author} {\bibinfo {author} {\bibfnamefont {M.}~\bibnamefont
  {Benito}}, \bibinfo {author} {\bibfnamefont {A.}~\bibnamefont
  {G{\'o}mez-Le{\'o}n}}, \bibinfo {author} {\bibfnamefont {V.}~\bibnamefont
  {Bastidas}}, \bibinfo {author} {\bibfnamefont {T.}~\bibnamefont {Brandes}}, \
  and\ \bibinfo {author} {\bibfnamefont {G.}~\bibnamefont {Platero}},\ }\href
  {\doibase 10.1103/PhysRevB.90.205127} {\bibfield  {journal} {\bibinfo
  {journal} {Phys. Rev. B}\ }\textbf {\bibinfo {volume} {90}},\ \bibinfo
  {pages} {205127} (\bibinfo {year} {2014})}\BibitemShut {NoStop}%
\bibitem [{Note3()}]{Note3}%
  \BibitemOpen
  \bibinfo {note} {For simplicity of language, here and in the following we
  employ the term ``condensate'' although, strictly speaking, in one-dimension
  only a ``quasi-condensate'' exists.}\BibitemShut {Stop}%
\bibitem [{\citenamefont {Mueller}\ \emph {et~al.}(2006)\citenamefont
  {Mueller}, \citenamefont {Ho}, \citenamefont {Ueda},\ and\ \citenamefont
  {Baym}}]{mueller}%
  \BibitemOpen
  \bibfield  {author} {\bibinfo {author} {\bibfnamefont {E.~J.}\ \bibnamefont
  {Mueller}}, \bibinfo {author} {\bibfnamefont {T.-L.}\ \bibnamefont {Ho}},
  \bibinfo {author} {\bibfnamefont {M.}~\bibnamefont {Ueda}}, \ and\ \bibinfo
  {author} {\bibfnamefont {G.}~\bibnamefont {Baym}},\ }\href {\doibase
  10.1103/PhysRevA.74.033612} {\bibfield  {journal} {\bibinfo  {journal} {Phys.
  Rev. A}\ }\textbf {\bibinfo {volume} {74}},\ \bibinfo {pages} {033612}
  (\bibinfo {year} {2006})}\BibitemShut {NoStop}%
\bibitem [{\citenamefont {Di~Liberto}\ \emph {et~al.}(2011)\citenamefont
  {Di~Liberto}, \citenamefont {Tieleman}, \citenamefont {Branchina},\ and\
  \citenamefont {Smith}}]{marco}%
  \BibitemOpen
  \bibfield  {author} {\bibinfo {author} {\bibfnamefont {M.}~\bibnamefont
  {Di~Liberto}}, \bibinfo {author} {\bibfnamefont {O.}~\bibnamefont
  {Tieleman}}, \bibinfo {author} {\bibfnamefont {V.}~\bibnamefont {Branchina}},
  \ and\ \bibinfo {author} {\bibfnamefont {C.~M.}\ \bibnamefont {Smith}},\
  }\href {\doibase 10.1103/PhysRevA.84.013607} {\bibfield  {journal} {\bibinfo
  {journal} {Phys. Rev. A}\ }\textbf {\bibinfo {volume} {84}},\ \bibinfo
  {pages} {013607} (\bibinfo {year} {2011})}\BibitemShut {NoStop}%
\bibitem [{\citenamefont {Sowi\ifmmode~\acute{n}\else
  \'{n}\fi{}ski}(2012)}]{sowinski}%
  \BibitemOpen
  \bibfield  {author} {\bibinfo {author} {\bibfnamefont {T.}~\bibnamefont
  {Sowi\ifmmode~\acute{n}\else \'{n}\fi{}ski}},\ }\href {\doibase
  10.1103/PhysRevA.85.065601} {\bibfield  {journal} {\bibinfo  {journal} {Phys.
  Rev. A}\ }\textbf {\bibinfo {volume} {85}},\ \bibinfo {pages} {065601}
  (\bibinfo {year} {2012})}\BibitemShut {NoStop}%
\bibitem [{\citenamefont {Cazalilla}\ \emph {et~al.}(2011)\citenamefont
  {Cazalilla}, \citenamefont {Citro}, \citenamefont {Giamarchi}, \citenamefont
  {Orignac},\ and\ \citenamefont {Rigol}}]{cazalilla}%
  \BibitemOpen
  \bibfield  {author} {\bibinfo {author} {\bibfnamefont {M.}~\bibnamefont
  {Cazalilla}}, \bibinfo {author} {\bibfnamefont {R.}~\bibnamefont {Citro}},
  \bibinfo {author} {\bibfnamefont {T.}~\bibnamefont {Giamarchi}}, \bibinfo
  {author} {\bibfnamefont {E.}~\bibnamefont {Orignac}}, \ and\ \bibinfo
  {author} {\bibfnamefont {M.}~\bibnamefont {Rigol}},\ }\href {\doibase
  10.1103/RevModPhys.83.1405} {\bibfield  {journal} {\bibinfo  {journal} {Rev.
  Mod. Phys.}\ }\textbf {\bibinfo {volume} {83}},\ \bibinfo {pages} {1405}
  (\bibinfo {year} {2011})}\BibitemShut {NoStop}%
\bibitem [{\citenamefont {Tomonaga}(1950)}]{tomonaga}%
  \BibitemOpen
  \bibfield  {author} {\bibinfo {author} {\bibfnamefont {S.}~\bibnamefont
  {Tomonaga}},\ }\href {https://doi.org/10.1143/ptp/5.4.544} {\bibfield
  {journal} {\bibinfo  {journal} {Progress of Theoretical Physics}\ }\textbf
  {\bibinfo {volume} {5}},\ \bibinfo {pages} {544} (\bibinfo {year}
  {1950})}\BibitemShut {NoStop}%
\bibitem [{\citenamefont {Luttinger}(1963)}]{luttinger}%
  \BibitemOpen
  \bibfield  {author} {\bibinfo {author} {\bibfnamefont {J.~M.}\ \bibnamefont
  {Luttinger}},\ }\href {\doibase 10.1063/1.1704046} {\bibfield  {journal}
  {\bibinfo  {journal} {Journal of Mathematical Physics}\ }\textbf {\bibinfo
  {volume} {4}},\ \bibinfo {pages} {1154} (\bibinfo {year} {1963})}\BibitemShut
  {NoStop}%
\bibitem [{\citenamefont {Haldane}(1981)}]{haldane}%
  \BibitemOpen
  \bibfield  {author} {\bibinfo {author} {\bibfnamefont {F.~D.~M.}\
  \bibnamefont {Haldane}},\ }\href {\doibase 10.1103/PhysRevLett.47.1840}
  {\bibfield  {journal} {\bibinfo  {journal} {Phys. Rev. Lett.}\ }\textbf
  {\bibinfo {volume} {47}},\ \bibinfo {pages} {1840} (\bibinfo {year}
  {1981})}\BibitemShut {NoStop}%
\bibitem [{\citenamefont {K{\"u}hner}, \citenamefont {White},\ and\
  \citenamefont {Monien}(2000)}]{white}%
  \BibitemOpen
  \bibfield  {author} {\bibinfo {author} {\bibfnamefont {T.~D.}\ \bibnamefont
  {K{\"u}hner}}, \bibinfo {author} {\bibfnamefont {S.~R.}\ \bibnamefont
  {White}}, \ and\ \bibinfo {author} {\bibfnamefont {H.}~\bibnamefont
  {Monien}},\ }\href {\doibase 10.1103/PhysRevB.61.12474} {\bibfield  {journal}
  {\bibinfo  {journal} {Phys. Rev. B}\ }\textbf {\bibinfo {volume} {61}},\
  \bibinfo {pages} {12474} (\bibinfo {year} {2000})}\BibitemShut {NoStop}%
\bibitem [{\citenamefont {Ejima}, \citenamefont {Fehske},\ and\ \citenamefont
  {Gebhard}(2011)}]{gebhard}%
  \BibitemOpen
  \bibfield  {author} {\bibinfo {author} {\bibfnamefont {S.}~\bibnamefont
  {Ejima}}, \bibinfo {author} {\bibfnamefont {H.}~\bibnamefont {Fehske}}, \
  and\ \bibinfo {author} {\bibfnamefont {F.}~\bibnamefont {Gebhard}},\ }\href
  {\doibase 10.1209/0295-5075/93/30002} {\bibfield  {journal} {\bibinfo
  {journal} {EPL}\ }\textbf {\bibinfo {volume} {93}},\ \bibinfo {pages} {30002}
  (\bibinfo {year} {2011})}\BibitemShut {NoStop}%
\bibitem [{\citenamefont {Giamarchi}(2003)}]{giamarchi}%
  \BibitemOpen
  \bibfield  {author} {\bibinfo {author} {\bibfnamefont {T.}~\bibnamefont
  {Giamarchi}},\ }\href {https://books.google.co.uk/books?id=GVeuKZLGMZ0C}
  {\emph {\bibinfo {title} {Quantum Physics in One Dimension}}},\ International
  Series of Monographs on Physics\ (\bibinfo  {publisher} {Clarendon Press},\
  \bibinfo {year} {2003})\BibitemShut {NoStop}%
\bibitem [{Note4()}]{Note4}%
  \BibitemOpen
  \bibinfo {note} {By independent Luttinger liquids, we mean here two sets of
  bosons $A$ and $B$, occupying different regions of the one-particle Hilbert
  space (here around momenta $\pm \pi /2$) such that the total density
  fluctuation operator can be written as $\delta n_A (x) + \delta n_B (x)$
  while ${\delimiter "426830A \Psi | \delta n_A (x) \delta n_B (x') | \Psi
  \delimiter "526930B } = 0$.}\BibitemShut {Stop}%
\bibitem [{Note5()}]{Note5}%
  \BibitemOpen
  \bibinfo {note} {Umklapp processes with final momenta $(p-k,q+k+2\pi )$ yield
  the same $F$ value.}\BibitemShut {Stop}%
\bibitem [{\citenamefont {Novi\ifmmode~\check{c}\else \v{c}\fi{}enko},
  \citenamefont {Anisimovas},\ and\ \citenamefont {Juzeli\ifmmode~\bar{u}\else
  \={u}\fi{}nas}(2017)}]{parametric_floquet}%
  \BibitemOpen
  \bibfield  {author} {\bibinfo {author} {\bibfnamefont {V.}~\bibnamefont
  {Novi\ifmmode~\check{c}\else \v{c}\fi{}enko}}, \bibinfo {author}
  {\bibfnamefont {E.}~\bibnamefont {Anisimovas}}, \ and\ \bibinfo {author}
  {\bibfnamefont {G.}~\bibnamefont {Juzeli\ifmmode~\bar{u}\else
  \={u}\fi{}nas}},\ }\href {\doibase 10.1103/PhysRevA.95.023615} {\bibfield
  {journal} {\bibinfo  {journal} {Phys. Rev. A}\ }\textbf {\bibinfo {volume}
  {95}},\ \bibinfo {pages} {023615} (\bibinfo {year} {2017})}\BibitemShut
  {NoStop}%
\bibitem [{\citenamefont {{Reitter}}\ \emph {et~al.}(2017)\citenamefont
  {{Reitter}}, \citenamefont {{N{\"a}ger}}, \citenamefont {{Wintersperger}},
  \citenamefont {{Str{\"a}ter}}, \citenamefont {{Bloch}}, \citenamefont
  {{Eckardt}},\ and\ \citenamefont {{Schneider}}}]{bloch_potassium}%
  \BibitemOpen
  \bibfield  {author} {\bibinfo {author} {\bibfnamefont {M.}~\bibnamefont
  {{Reitter}}}, \bibinfo {author} {\bibfnamefont {J.}~\bibnamefont
  {{N{\"a}ger}}}, \bibinfo {author} {\bibfnamefont {K.}~\bibnamefont
  {{Wintersperger}}}, \bibinfo {author} {\bibfnamefont {C.}~\bibnamefont
  {{Str{\"a}ter}}}, \bibinfo {author} {\bibfnamefont {I.}~\bibnamefont
  {{Bloch}}}, \bibinfo {author} {\bibfnamefont {A.}~\bibnamefont {{Eckardt}}},
  \ and\ \bibinfo {author} {\bibfnamefont {U.}~\bibnamefont {{Schneider}}},\
  }\href@noop {} {\bibfield  {journal} {\bibinfo  {journal} {ArXiv e-prints}\ }
  (\bibinfo {year} {2017})},\ \Eprint {http://arxiv.org/abs/1706.04819}
  {arXiv:1706.04819 [cond-mat.quant-gas]} \BibitemShut {NoStop}%
\bibitem [{\citenamefont {Abanin}, \citenamefont {De~Roeck},\ and\
  \citenamefont {Huveneers}(2015)}]{abanin_2015}%
  \BibitemOpen
  \bibfield  {author} {\bibinfo {author} {\bibfnamefont {D.~A.}\ \bibnamefont
  {Abanin}}, \bibinfo {author} {\bibfnamefont {W.}~\bibnamefont {De~Roeck}}, \
  and\ \bibinfo {author} {\bibfnamefont {F.~m.~c.}\ \bibnamefont {Huveneers}},\
  }\href {\doibase 10.1103/PhysRevLett.115.256803} {\bibfield  {journal}
  {\bibinfo  {journal} {Phys. Rev. Lett.}\ }\textbf {\bibinfo {volume} {115}},\
  \bibinfo {pages} {256803} (\bibinfo {year} {2015})}\BibitemShut {NoStop}%
\bibitem [{\citenamefont {Abanin}\ \emph {et~al.}(2017)\citenamefont {Abanin},
  \citenamefont {De~Roeck}, \citenamefont {Ho},\ and\ \citenamefont
  {Huveneers}}]{abanin_2017}%
  \BibitemOpen
  \bibfield  {author} {\bibinfo {author} {\bibfnamefont {D.~A.}\ \bibnamefont
  {Abanin}}, \bibinfo {author} {\bibfnamefont {W.}~\bibnamefont {De~Roeck}},
  \bibinfo {author} {\bibfnamefont {W.~W.}\ \bibnamefont {Ho}}, \ and\ \bibinfo
  {author} {\bibfnamefont {F.~m.~c.}\ \bibnamefont {Huveneers}},\ }\href
  {\doibase 10.1103/PhysRevB.95.014112} {\bibfield  {journal} {\bibinfo
  {journal} {Phys. Rev. B}\ }\textbf {\bibinfo {volume} {95}},\ \bibinfo
  {pages} {014112} (\bibinfo {year} {2017})}\BibitemShut {NoStop}%
\bibitem [{\citenamefont {Bukov}\ \emph {et~al.}(2015)\citenamefont {Bukov},
  \citenamefont {Gopalakrishnan}, \citenamefont {Knap},\ and\ \citenamefont
  {Demler}}]{bukov_2015}%
  \BibitemOpen
  \bibfield  {author} {\bibinfo {author} {\bibfnamefont {M.}~\bibnamefont
  {Bukov}}, \bibinfo {author} {\bibfnamefont {S.}~\bibnamefont
  {Gopalakrishnan}}, \bibinfo {author} {\bibfnamefont {M.}~\bibnamefont
  {Knap}}, \ and\ \bibinfo {author} {\bibfnamefont {E.}~\bibnamefont
  {Demler}},\ }\href {\doibase 10.1103/PhysRevLett.115.205301} {\bibfield
  {journal} {\bibinfo  {journal} {Phys. Rev. Lett.}\ }\textbf {\bibinfo
  {volume} {115}},\ \bibinfo {pages} {205301} (\bibinfo {year}
  {2015})}\BibitemShut {NoStop}%
\bibitem [{\citenamefont {Rigol}\ and\ \citenamefont
  {Muramatsu}(2004)}]{muramatsu2004}%
  \BibitemOpen
  \bibfield  {author} {\bibinfo {author} {\bibfnamefont {M.}~\bibnamefont
  {Rigol}}\ and\ \bibinfo {author} {\bibfnamefont {A.}~\bibnamefont
  {Muramatsu}},\ }\href {\doibase 10.1103/PhysRevLett.93.230404} {\bibfield
  {journal} {\bibinfo  {journal} {Phys. Rev. Lett.}\ }\textbf {\bibinfo
  {volume} {93}},\ \bibinfo {pages} {230404} (\bibinfo {year}
  {2004})}\BibitemShut {NoStop}%
\bibitem [{\citenamefont {Rodriguez}\ \emph {et~al.}(2006)\citenamefont
  {Rodriguez}, \citenamefont {Manmana}, \citenamefont {Rigol}, \citenamefont
  {Noack},\ and\ \citenamefont {Muramatsu}}]{muramatsu2006}%
  \BibitemOpen
  \bibfield  {author} {\bibinfo {author} {\bibfnamefont {K.}~\bibnamefont
  {Rodriguez}}, \bibinfo {author} {\bibfnamefont {S.}~\bibnamefont {Manmana}},
  \bibinfo {author} {\bibfnamefont {M.}~\bibnamefont {Rigol}}, \bibinfo
  {author} {\bibfnamefont {R.}~\bibnamefont {Noack}}, \ and\ \bibinfo {author}
  {\bibfnamefont {A.}~\bibnamefont {Muramatsu}},\ }\href {\doibase
  10.1088/1367-2630/8/8/169} {\bibfield  {journal} {\bibinfo  {journal} {New
  Journal of Physics}\ }\textbf {\bibinfo {volume} {8}},\ \bibinfo {pages}
  {169} (\bibinfo {year} {2006})}\BibitemShut {NoStop}%
\bibitem [{\citenamefont {Vidmar}\ \emph {et~al.}(2015)\citenamefont {Vidmar},
  \citenamefont {Ronzheimer}, \citenamefont {Schreiber}, \citenamefont {Braun},
  \citenamefont {Hodgman}, \citenamefont {Langer}, \citenamefont
  {Heidrich-Meisner}, \citenamefont {Bloch},\ and\ \citenamefont
  {Schneider}}]{schneider}%
  \BibitemOpen
  \bibfield  {author} {\bibinfo {author} {\bibfnamefont {L.}~\bibnamefont
  {Vidmar}}, \bibinfo {author} {\bibfnamefont {J.~P.}\ \bibnamefont
  {Ronzheimer}}, \bibinfo {author} {\bibfnamefont {M.}~\bibnamefont
  {Schreiber}}, \bibinfo {author} {\bibfnamefont {S.}~\bibnamefont {Braun}},
  \bibinfo {author} {\bibfnamefont {S.~S.}\ \bibnamefont {Hodgman}}, \bibinfo
  {author} {\bibfnamefont {S.}~\bibnamefont {Langer}}, \bibinfo {author}
  {\bibfnamefont {F.}~\bibnamefont {Heidrich-Meisner}}, \bibinfo {author}
  {\bibfnamefont {I.}~\bibnamefont {Bloch}}, \ and\ \bibinfo {author}
  {\bibfnamefont {U.}~\bibnamefont {Schneider}},\ }\href {\doibase
  10.1103/PhysRevLett.115.175301} {\bibfield  {journal} {\bibinfo  {journal}
  {Phys. Rev. Lett.}\ }\textbf {\bibinfo {volume} {115}},\ \bibinfo {pages}
  {175301} (\bibinfo {year} {2015})}\BibitemShut {NoStop}%
\end{thebibliography}%

\appendix

\section{\label{app_eff} Effective Hamiltonian}

To obtain the effective description of the system [Eq. \eqref{eq:driven_hubb_ham}] it is useful to work in the momentum representation. We employ the discrete Fourier transform of the creation and annihilation operators
\begin{align}
& a^\dagger_x = \frac{1}{\sqrt{N}}\sum^{N-1}_{l=0} e^{- \im k_l x}a^\dagger_{k_l}~,
\label{eq:fourier_into_mom_space}
\end{align}
where $k_l = 2 \pi l/N $, with the corresponding
inverse transform
\begin{align}
& a^\dagger_{k_l} = \frac{1}{\sqrt{N}}\sum^{N-1}_{x=0} e^{\im k_l x}a^\dagger_{x}~.
\label{eq:fourier_into_pos_space}
\end{align}
The Hamiltonian then becomes
\begin{align}
\begin{aligned}
H(t) & =  -2 \Jac \cos (\omega t)\sum_{l = 0}^{N-1}  \cos( k_l ) a^\dagger_{k_l}a_{k_l}
\\
&+ \frac{U}{2 N} \sum^{N-1}_{l,m,n,p = 0}  \delta_{k_l+k_m,k_n+k_p} a^\dagger_{k_p}a^\dagger_{k_n}a_{k_m}a_{k_l}~.
\end{aligned}
\label{eq:hub_ham_momentum}
\end{align}
Transforming into the interaction picture helps to find the non-vanishing
time-averaged contributions to the effective Hamiltonian. It is straightforward to see that in Eq. \eqref{eq:driven_hubb_ham} the hopping immediately vanishes when averaged over one period. Using the unitary transformation
\begin{equation}
R(t)= \exp \left(- 2 \im \frac{ J_{\ac}}{\omega}\sin (\omega t) \sum^{N-1}_{k_l=0} \cos\left(k_l\right)a^\dagger_{k_l}a_{k_l}  \right)~,
\label{eq:int_pic_trafo}
\end{equation}
takes Eq. \eqref{eq:hub_ham_momentum} into
\begin{equation}
\tilde{H}(t) = \im \dot{R}(t)R^\dagger(t)+ R(t) H(t) R^\dagger(t) ~.
\label{eq:abstr_Ham_trafo}
\end{equation}
In performing the transformation it is essential to rearrange terms of the form
\begin{align}
&\begin{aligned}
f(\lambda):= &e^{- \im \lambda \cos\left( k_u \right) a^\dagger_{k_u}a_{k_u}}
\\
&\hspace{.7cm}\times\left(
a^\dagger_{k_p}a^\dagger_{k_n}a_{k_m}a_{k_l}
\right)
e^{\im \lambda \cos\left( k_u \right) a^\dagger_{k_u}a_{k_u}}
\end{aligned}
\\
&\hphantom{f(\lambda)} = U(\lambda)  a^\dagger_{k_p}a^\dagger_{k_n}a_{k_m}a_{k_l} U^\dagger (\lambda) ~.
\end{align}
Differentiation with respect to $\lambda$ results in
\begin{align}
\frac{\d f (\lambda)}{\d \lambda} &= \im \cos ( k_u ) U(\lambda) \left[ a^\dagger_{k_p}a^\dagger_{k_n}a_{k_m}a_{k_l}, a^\dagger_{k_u}a_{k_u} \right]  U^\dagger (\lambda)
\\
& =
\im \cos \left( k_u \right)\left(\delta_{lu}+\delta_{mu}-\delta_{nu}-\delta_{pu}\right)f(\lambda)~,
\end{align}
which implies that
\begin{equation}
\quad f(\lambda) = e^{\im \lambda \cos\left( k_u \right)\left(\delta_{lu}+\delta_{mu}-\delta_{nu}-\delta_{pu}\right)}a^\dagger_{k_p}a^\dagger_{k_n}a_{k_m}a_{k_l}~.
\end{equation}
Using the last equation in combination with Eq. \eqref{eq:abstr_Ham_trafo} and Eq. \eqref{eq:int_pic_trafo} gives
\begin{align}
&\begin{aligned}
\tilde{H}(t) & = \frac{U}{2 N} \sum^{N-1}_{l,m,n,p = 0} \delta_{k_l+k_m,k_n+k_p}
\\
& \times
e^{ 2 \im \frac{ J_{\ac}}{\omega}\sin (\omega t)[\cos(k_l)+\cos(k_m)-\cos(k_n)-\cos(k_p)]}
\\&\hspace{4cm}\times a^\dagger_{k_p}a^\dagger_{k_n}a_{k_m}a_{k_l}~.
\end{aligned}
\end{align}
Time averaging then provides us with the effective Hamiltonian
\begin{widetext}
\begin{eqnarray}
\Heff &=& \frac{1}{T} \int_0^T {\rm d}t\, \tilde{H}(t) \nonumber \\
&=& \frac{U}{2 N} \sum^{N-1}_{l,m,n,p = 0} \delta_{k_l+k_m,k_n+k_p}
\mathcal{J}_0 \left\{ 2\kappa \left[ \cos(k_l)+\cos(k_m)-\cos(k_n)-\cos(k_p) \right] \right\}
a^\dagger_{k_p}a^\dagger_{k_n}a_{k_m}a_{k_l}~,
\label{eq:Heff_mom_rep}
\end{eqnarray}
%
%\label{eq:Heff_mom_rep}
\end{widetext}
where $\mathcal{J}_0$ is the Bessel function of zeroth order and $\kappa = J_{\rm ac}/\omega$.
Using the inverse transformation in Eq. \eqref{eq:fourier_into_pos_space} we find in position space
\begin{align}
\begin{aligned}
\Heff =\sum_{w,x,y,z=0}^{N-1} Q_{wxyz}  a^\dagger_w a^\dagger_x a_y a_z~,
\end{aligned}
\label{eq:ham_eff_mom}
\end{align}
where
\begin{align}
\begin{aligned}
Q_{wxyz} & =  \frac{U}{2N^3}\sum_{l,m,n=0}^{N-1} e^{\im [k_l(x-z)+k_m(x-y)-k_n(x-w)]}
\\
&\hspace{-.5cm} \times \mathcal{J}_0\left\{ 2 \kappa \left[\cos(k_l)+\cos(k_m)
\right.\right. \\ &\hspace{1.2cm}\left.\left.-\cos(k_l+k_m-k_n)-\cos(k_n)\right]\right\}~,
\end{aligned}
\label{q_amp}
\end{align}
which is the form of the effective Hamiltonian given in Eqs. (\ref{eq:ham_eff}
-\ref{M-kappa}).

\section{\label{app_dip} Dipole approximation}

\begin{figure}
\begin{center}
\includegraphics[width=0.40\textwidth,clip=true]{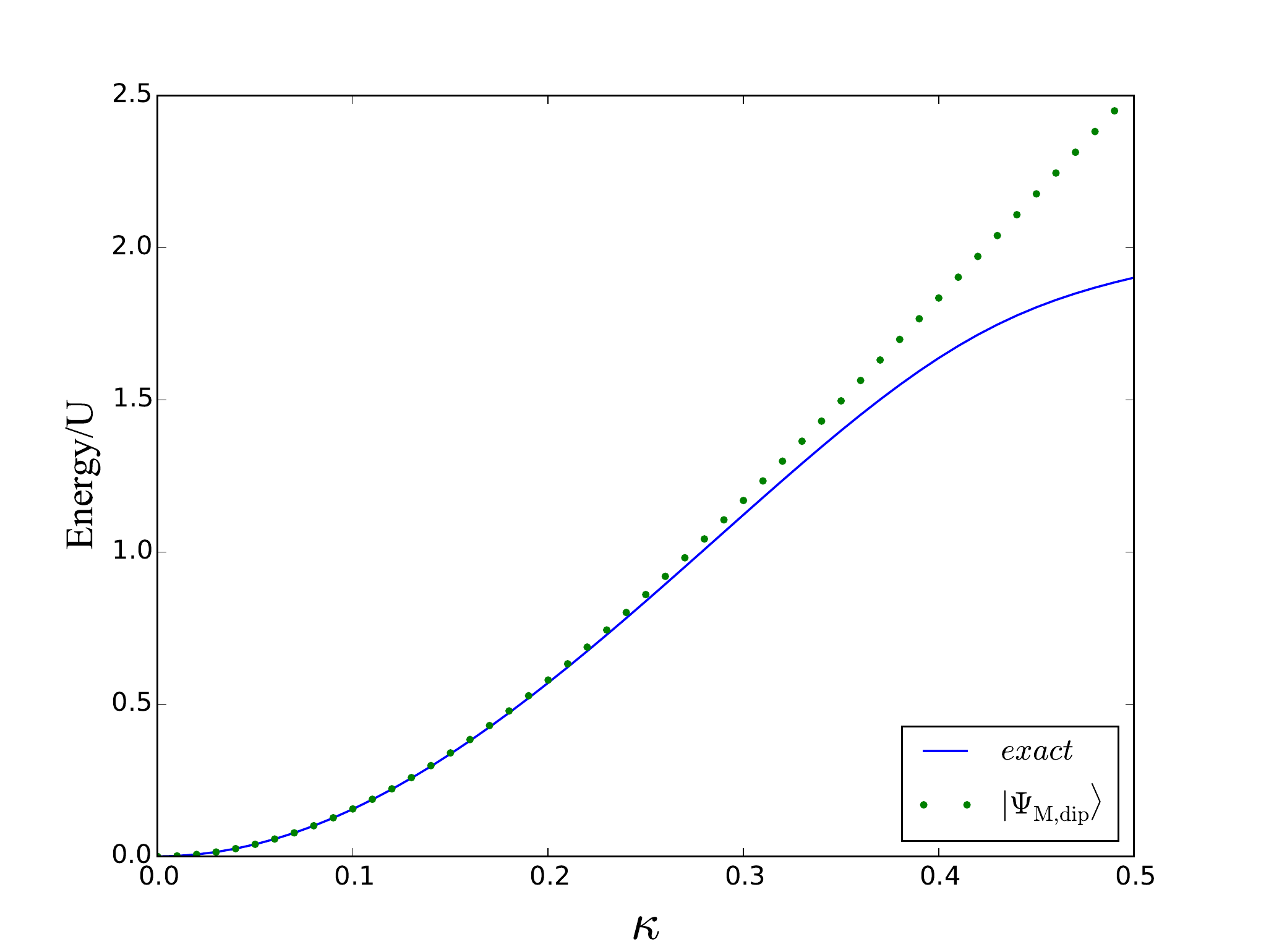}
\end{center}
\caption{The exact ground state energy of the effective Hamiltonian is approximated well by the dipole approximation until $\kappa \simeq 0.3$. Beyond this value, other excitations of the undriven Hamiltonian start to contribute significantly to the ground state.}
\label{fig:dip_ener_comp}
\end{figure}

Using the effective Hamiltonian Eq. \eqref{eq:ham_eff}, we can calculate
the expectation value of the energy of the dipole state \eqref{eq:ansatz}
and then vary its parameters ($\alpha$ and $\beta$) to minimize this energy.
This then provides a variational approximation for the ground state of the
system. As the normalization of the state requires $\alpha^2 + \beta^2 = 1$,
we thus wish to minimize $\langle H_{\rm eff}\rangle_\alpha$, 
where $\beta = \sqrt{1-\alpha^2}$.
The variational energy can then be written as
\begin{equation}
\langle H_{\rm eff}\rangle_\alpha = \alpha^2(A-C) + \alpha\sqrt{1-\alpha^2}B+C~,
\end{equation}
where
\begin{widetext}
\begin{align}
&A = \langle {\rm Mott}|H_{\rm eff}|{\rm Mott}\rangle = 2 \sum_{x\neq y} Q_{xyyx}\\
&B =2 \langle {\rm Mott}|H_{\rm eff}|\Psi_{\rm dip}\rangle = 8 \sqrt{N}\Big(\sum_{x\neq0}+\sum_{x\neq 0, 2}\Big)Q_{2xx0}
\\
&\begin{aligned}
C = \langle \Psi_{\rm dip}| H_{\rm eff}|\Psi_{\rm dip}\rangle =&
2 \sum_{x\neq y} Q_{xyyx}+4\sum_{x\neq 0, 2, 4}Q_{4xx0}
+ 4\sum_{x\neq 0, 2, 4}Q_{4,x,x+2,2}
-4\sum_{x\neq 0, 2}Q_{0xx0}
\\
&+\Big(2\sum_{\substack{x\neq 2\\ x\neq N-2}}+6\sum_{\substack{x\neq 0, 2\\ x\neq N-2}}\Big)Q_{2,x+2,x,0} +
\Big(8\sum_{x\neq 0}-4\sum_{x\neq 2}\Big)Q_{2xx2}\\
&+2 \Big(\sum_{x\neq 0}+\sum_{x\neq 0, 2}+\sum_{\substack{x\neq 0\\x\neq N-2}}+\sum_{\substack{x\neq 0, 2\\x\neq N-2}}\Big)Q_{2xx,N-2}+ 8 Q_{4220} -4Q_{2442}-6Q_{2222} ~.
\end{aligned}
\end{align}
\end{widetext}
The energy is minimized for
\begin{align}
\alpha = \left\{\begin{aligned}
&\sqrt{\frac{1}{2}+\frac{1}{2}\left(1+\frac{2 B^2}{(A-C)^2}\right)^{-\frac{1}{2}}}~\mathrm{for}~ B < 0\\
- \ &\sqrt{\frac{1}{2}+\frac{1}{2}\left(1+\frac{2 B^2}{(A-C)^2}\right)^{-\frac{1}{2}}}~\mathrm{for}~ B > 0
\end{aligned} \right.
\end{align}
The dipole approximation works well for $\kappa < 0.25$, as can be seen in Fig. \ref{fig:dip_ener_comp}.
Even though the ansatz is not too far from the exact energy even for $\kappa = 0.4$, it is unable to describe the phase transition since its density-density correlation
\begin{equation}
N(r) = \frac{1}{N}(1-\alpha^2)(2 \delta_0 -\delta_{r,2}-\delta_{r,-2})
\end{equation}
is always short ranged. This means that across the transition, higher-order excitations of the undriven Hamiltonian are essential to produce the long range density correlations,
indicating a transition to a superfluid.

\section{\label{app_luttinger} Two-component Luttinger liquids}
Let us assume we have two different Luttinger liquids (call them A and B) that coexist in the same region of space but which are independent in terms of their density fluctuations
\begin{equation}
\langle \delta n_A(x) \delta n_B(x')\rangle = 0\, ,\label{eq:ABcomm} 
\end{equation}
where 
\begin{equation}
\delta n_{\alpha}(x) \equiv n_{\alpha}(x) - \langle n_{\alpha}(x)\rangle \,
\end{equation}
and $\alpha=A,B$.
Let us also assume that we can only measure the total density
\begin{equation}
n(x) = n_A(x) + n_B(x) \, .
\label{eq:total-density}
\end{equation}
Since A and B are both Luttinger liquids, they separately satisfy Eq. \eqref{decay} 
\begin{align}
N_\alpha(r\rightarrow\infty) &=  \frac{1}{L} \sum_{x} \langle\delta \tilde{n}_{\alpha}(x)\delta \tilde{n}_{\alpha}(x+r)\rangle_\alpha  
\\&= -\frac{1}{2 K_b (\pi r)^2}\, ,\label{eq:Nr_Kb}
\end{align}
where 
\begin{equation}
\delta \tilde{n}_\alpha \equiv n_\alpha(x) - \langle n_\alpha(x) \rangle_\alpha
\label{eq:av_fluc_total_state}
\end{equation} 
and the average $\langle \cdot \rangle_\alpha$ is taken with respect to a state that only contains species $\alpha$. Here, $L$ is the number of sites and we assume $K_b < 1$ is identical for the two liquids.
We can apply the theory of \cite{cazalilla} and note that, for the commensurate case, $K_b = 1/2$ signals the onset of superfluidity for both A and B. Importantly, $K_b$ only depends on the commensurability of particles and sites and not on the average density. 
We expect the total density fluctuations to follow a similar law,  
\begin{equation}
N(r \rightarrow \infty) = \frac{1}{L} \sum_x \langle \delta n(x) \delta n(x+r)\rangle  = - \frac{1}{2Q(\pi r)^2} \, ,
\label{Nr-Q-single}
\end{equation} 
and address the question of how $Q$ relates to $K_b$.
Eqs. \eqref{eq:ABcomm} - \eqref{eq:total-density} imply  
\begin{align}
&\begin{aligned}
N(r\rightarrow\infty) =& \frac{1}{L} \sum_x  \Big[\langle\delta n_A(x) \delta n_A(x + r)\rangle \\&\hspace{1cm}+ \langle\delta n_B(x) \delta n_B(x+r)\rangle\Big] \, 
\end{aligned}\label{eq:nr_lutt_decoupled1}\\
&\begin{aligned}
\hphantom{N(r\rightarrow\infty)} =& \frac{1}{2L} \sum_x  \Big[\langle\delta \tilde{n}_A(x) \delta \tilde{n}_A(x + r)\rangle_A \\&\hspace{1cm}+ \langle\delta \tilde{n}_B(x) \delta \tilde{n}_B(x+r)\rangle_B\Big] \, ,
\end{aligned}
\label{eq:nr_lutt_decoupled2}
\end{align}
where one assumes that the joint system of A and B is in a state such that Eq. \eqref{eq:nr_lutt_decoupled2} is satisfied. Using Eq. \eqref{eq:Nr_Kb} one can readily conclude that $K_b = Q$. 

Now let us see whether the above assumptions apply to a many-body state of cat type
\begin{equation}
|\Psi \rangle = \frac{1}{\sqrt{2}} \left( A^{\dagger} +  B^{\dagger} \right) |\text{vac} \rangle \, ,
\label{cat-creation}
\end{equation}
with $ |\Phi_A\rangle \equiv A^\dagger |{\rm vac} \rangle $ and $|\Phi_B\rangle \equiv B^\dagger |{\rm vac} \rangle$ being orthonormal $ \langle \Phi_\alpha|\Phi_{\alpha'}\rangle = \delta_{\alpha,\alpha'}$.  
Since $A^\dagger$ and $B^\dagger$ are, by construction, operators that only create A- and B-type bosons, respectively, we can write   
\begin{equation}
n_A(x) B^{\dagger} |\text{vac} \rangle = 0, \; \; 
n_B(x) A^{\dagger} |\text{vac} \rangle = 0 \, .
\label{eq:LLind2}
\end{equation}
\\
Clearly the cat state \eqref{cat-creation} satisfies \eqref{eq:nr_lutt_decoupled2}, with 
$\langle O \rangle_\alpha = \langle\Phi_\alpha| O |\Phi_\alpha \rangle$. Thus we infer that the result $Q = K_b$ also applies to the above cat state. 

It only remains to analyze to what extent our fragmented-condensate boson system conforms to the two Luttinger-liquid picture defined above. For that purpose we focus on whether our fragmented condensate satisfies Eqs. \eqref{cat-creation} and \eqref{eq:LLind2}. 
Numerical evidence discussed in Section \ref{2-particle-momentum} strongly suggests that our ground state is of the cat type \eqref{cat-creation}, where $A$, $B$ are operators that create many bosons peaked around $\pm \pi/2$, respectively.
We now need to justify whether this constitutes two independent boson species in the sense of Eqs. \eqref{eq:ABcomm} and \eqref{eq:total-density}. 
We focus on the homogeneous case for which the average density is position independent. 
In general, if we define
\begin{equation}
\delta \hat{n}(x) \equiv \hat{n}(x)- \frac{1}{L} \sum_x \hat{n}(x) \ , 
\label{eq:av_dens_op}
\end{equation}
we can write 
\begin{equation}
\delta \hat{n}(x) = \frac{1}{L} \sum_k \sum_{q\neq 0} e^{iqx}a_k^{\dagger}a_{k+q} \, ,
\label{density-kq}
\end{equation}
where $a_k^{\dagger}$ creates a boson of momentum $k$. Given that we are only interested in the long-distance (large $r$) behavior of the correlation functions, we can assume that only small $q$ values will be relevant in \eqref{density-kq}. Then we can divide the operators \eqref{eq:av_dens_op} and \eqref{density-kq} into two contributions, labeled A [with $k \in (0,\pi)$] and B [with $k \in (-\pi,0)$], both restricted to small $q$. Thus, for instance,
\begin{equation}
\delta \hat{n}_A(x) = \frac{1}{L} \sum_{k \in (0,\pi)} \sum^{\rm small}_{q\neq 0} e^{iqx}a_k^{\dagger}a_{k+q} \, ,
\label{density-A}
\end{equation}
if we correspondingly define $\delta \hat{n}_A = \hat{n}_A - \frac{1}{L}\sum_x \hat{n}_A(x)$.
This allows us to approximately decompose the total density fluctuation operator as
\begin{equation}
\delta \hat{n}(x) \simeq \delta \hat{n}_A(x) + \delta \hat{n}_B(x) \, .
\label{density-AB}
\end{equation}
Clearly $\hat{n}_A(x)$ and $\hat{n}_B(x)$ satisfy Eq. \eqref{eq:LLind2}. This permits us to infer 
Eq. \eqref{eq:ABcomm}, which is crucial for the derivation of Eq. \eqref{eq:nr_lutt_decoupled1}.
Thus, the cat-type state in Eq. \eqref{cat-creation} with $A^\dagger$ and $B^\dagger$ creating bosons near $\pm \pi/2$  can rightly be viewed as formed by two independent condensates in the sense of Eqs. \eqref{eq:ABcomm} and \eqref{eq:total-density}. 
This provides a plausible explanation of why our peculiar one-dimensional system, involving two condensates that cannot be distinguished in terms of density, has the same coefficient $Q = K_b$ as the standard, one-component Luttinger liquid.

\end{document}